\documentclass{article}
\usepackage{makeidx}
\usepackage{amssymb}

\usepackage{amsthm}

\usepackage[section]{placeins}

  \usepackage[authoryear,round,longnamesfirst]{natbib}

\input{tcilatex}
\begin{document}

\sloppy

\title{Fair mixing: the case of{\ dichotomous preferences}}
\author{Haris Aziz$^{\ast }$, Anna Bogomolnaia$^{\ast \ast }$, and Herv\'{e}
Moulin$^{\ast \ast }$}
\date{ November 2017\\
$^{\ast }$ Data61, CSIRO and University of New South Wales\\
$^{\ast \ast }$ University of Glasgow and HSE St Petersburg}
\maketitle

\begin{abstract}
Agents vote to choose a fair mixture of public outcomes; each agent likes or
dislikes each outcome. We discuss three outstanding voting rules.

The \textit{Conditional Utilitarian }rule, a variant of the random dictator,
is Strategyproof and guarantees to any group of like-minded agents an
influence proportional to its size. It is easier to compute and more
efficient than the familiar \textit{Random Priority} rule. Its worst case
(resp. average) inefficiency is provably (resp. in numerical experiments) low
if the number of agents is low.

The efficient \textit{Egalitarian} rule protects similarly individual agents
but not coalitions. It is \textit{Excludable Strategyproof}: I do not want
to lie if I cannot consume outcomes I claim to dislike.

The efficient \textit{Nash Max Product }rule offers the strongest welfare
guarantees to coalitions, who can force any outcome with a probability
proportional to their size. But it fails even the excludable form of
Strategyproofness.
\end{abstract}

\section{Introduction}

Interactive democracy aka \textit{Liquid Democracy} (see e.g., %
\citep{Behr17a,Bri17a}) is a new approach to voting well suited for low stakes/
high frequency decisions, and easily implemented on the internet \citep{Gran17a}. An especially successful instance is budgetary participation %
\citep{Caba04a} where the stake-holders (citizens, employees of a firm, club
members) vote to decide which subset of public projects the community, firm,
or club should implement.

We discuss a stylized version of this process in the probabilistic voting
model~\citep{Fish84a,Gibb77a}. The guiding principle of our analysis is that
the selection of a single (deterministic) public outcome is prima facie
unfair: fairness requires compromise, we must select a \textit{mixture} of
several mutually exclusive outcomes. The mixture may come from actual
randomization, or the allocation of time-shares, or the distribution of a
fixed amount of some resource (e.g., money) over these outcomes. Some
typical examples follow.

In reference \citep{Caba04a}, the city authority must divide funds or staff
between several projects (library, sport center, concert hall) taking into
account the citizens' wishes. The scheduling of one or several weekly club
meetings (gym classes, chess club, study group) must accomodate the time
constraints reported by the club members. Or the local public TV, after
polling its audience, must divide broadcasting time between different
languages, or different types of program (news, sports, movies). In the 
\textit{fair knapsack} problem, the server schedules repeatedly jobs of
different reported or observed sizes under a capacity constraint, and must
pick a (random) serving protocol.

In all these examples, fairness requires to give some share of the public
resources to everyone: each club member should have access to some meetings;
everyone should enjoy at least some TV programs, etc.. This contrasts with
traditional high stakes/low frequency voting contexts, where the first best
is to select a single (deterministic) outcome, and randomization over
outcomes is only second best.\footnote{%
It is used\ to break ties, or to play the role of an absent deterministic
Condorcet winner: for instance \citep{LLL93b,ABBB15a,Bran17a} identifies a
lottery that, in a certain sense, wins the majority tournament.}

We run into the familiar conflict between protecting minorities and
submitting to the will of the majority~{\citep{Youn50a,Gord94a,PLSV00a}}. On
the one hand, the larger the support for a public outcome, the bigger should
be its share in the final compromise: numbers matter. On the other hand, we
must protect minorities with their idiosyncratic preferences for outcomes
disliked by the majority. So the club meetings will be more frequent when
many members can attend, but nobody will be entirely excluded; the knapsack
server will favor short jobs because this increases the number of satisfied
customers, but it cannot ignore long jobs entirely; and so on.

We analyze this tradeoff when preferences can be represented in a very
simple Facebook-style \textit{dichotomous }form: each agent likes or
dislikes each outcome, and her utility is simply the total share of her 
\textit{likes}. Agents in the knapsack problem care only about their
expected service time, and in the club example, about the number of meetings
they can attend. Though less natural in the public TV and the library
funding examples, where they rule out any complementarities between
outcomes, dichotomous preferences are still of practical interest because
they are easy to elicit.

We discuss the fairness and incentive compatibility properties of three
mostly well known social choice rules.\smallskip

\paragraph{\textit{Our results}.}

The \textit{Fair Share guarantee }principle is central to the fair division
literature since the earliest \textit{cake division} papers~\citep{Stei48a}.
In our model this is the \textit{Individual Fair Share} (IFS) axiom: each
one of the $n$ agents \textquotedblleft owns\textquotedblright\ a $1/n$-th
share of decision power, so she can ensure an outcome she likes at least $%
1/n $-th of the time (or with probability at least $1/n$). To capture the
more subtle ideas that minorities should be protected, and numbers should
matter as well, we strengthen IFS to \textit{Unanimous Fair Share} (UFS),
giving to any group of like-minded agents an influence proportional to its
size: so if 10\% of the agents have identical preferences they should like
the outcome at least 10\% of the time.

Our starting point is the impossibility result in \citep{BMS05a}, where our
model and the two fairness properties IFS and UFS appear first: \textit{no
mixing rule can be efficient, incentive compatible in the prior-free sense
of Strategyproofness (SP), and meet Unanimous, or even Individual, Fair Share%
}. We introduce new fairness and incentives properties and offer instead
possibility results. Three remarkable mixing rules (two of them well known)
meet IFS and achieve, loosely speaking, two out of the three goals of
efficiency, group fairness (in the sense of UFS or other more demanding
properties), and incentive compatibility.

Start with the \textit{Egalitarian} (EGAL) rule, adapting to our model a
celebrated principle of distributive justice. Taking the probability that
the selected outcome is liked by agent $i$ as her canonical utility, the
rule maximizes first the utility level we can guarantee to all agents; among
the corresponding mixtures, it maximizes the utility we can guarantee to all
agents but one; and so on. It is efficient and implements IFS, therefore it
is not strategyproof, by the above mentioned result. However if public
outcomes are non rival but \textit{excludable}, we can force agents to
consume only those outcomes they claim to like, so it becomes more costly to
fake a dislike and the strategyproofness is correspondingly weakened. A
meeting of the club is such an \textit{excludable} public outcome: it is
easy to exclude from the meeting those who reported they could not attend;
broadcasting via cable TV is similarly excludable, not so via aerial
broadcasting. The Egalitarian rule is efficient\ as well as \textit{%
Excludable Strategyproof }(EXSP): misreporting one's preferences does not
pay, provided an agent is excluded from consuming those public outcomes she
reportedly dislikes (Theorem 1). Thus weakening SP to EXSP resolves the
impossibility result.

But numbers do not matter to the egalitarian rule: it treats a unanimous
group of agents exactly as if it contained a single agent, so the UFS
property obviously fails. A related problem is that if I have a clone
(another agent with preferences identical to mine), I can simply stay home
and nothing will change. The \textit{Strict Participation }(PART$^{\ast }$)
axiom takes care of this disenfranchisement problem by insisting that
casting his vote is strictly benefitial to each voter. So the EGAL rule is
only appealing if we focus on individual guarantees and are comfortable
treating a homogenous group as a single person. This makes sense if the club
must offer some important training to its members. But in the budgetary
participation or the broadcasting examples, numbers should definitely matter.

The \textit{Conditional Utilitarian} (CUT)\textit{\ }rule is a simple
variant of the classic \textquotedblleft random dictator\textquotedblright .
Each agent identifies, among the outcomes he likes, those with the largest
support from the other agents: then he spreads the probability (time share)
of $1/n$ uniformly over the outcomes he likes. So the utilitarian concern is
conditional upon guaranteeing one's full utility first: \textit{charity
begins at home}. The CUT rule is related to, but much simpler than, the 
\textit{Random Priority }(RP) rule averaging outcomes of all deterministic
priority rules. Both rules are SP, meet PART$^{\ast }$ and guarantee UFS.
Therefore they are inefficient. But CUT is much easier to compute and
strictly more efficient than RP (Theorem 2). In numerical simulations
(Section 9) and for relatively small values of $n$, its inefficiency is
consistently low.

Our third rule is the familiar \textit{Nash Max Product }(NMP) rule picking
the mixture maximizing the product of individual utilities. It is efficient
and offers much stronger welfare guarantees to groups than UFS. We introduce
two requirements, each one a considerable strengthening of UFS, intuitively
but not logically related. The \textit{Core Fair Share} (CFS) property has
an incentive flavor in the spirit of cumulative voting~%
\citep{Gord94a,SaMa62a}: any group of agents can pool their shares of
decision power and object to the proposed mixture $z$ by enforcing another
mixture $z^{\prime }$ with a probability proportional to the group size.
Core Fair Share rules out any such objection. Finally \textit{Average Fair
Share} (AFS) applies to any coalition with a common liked outcome: the 
\textit{average} utility in such a group cannot be smaller than its relative
size. In simple examples, AFS limits very effectively the set of acceptable
efficient mixtures. Theorem 3 shows that the efficient NMP rule meets PART$%
^{\ast }$, CFS and AFS but fails even EXSP.

The results suggest several challenging open questions about the
impossibility frontiers of our model.

\section{Related literature}

Budgetary participation is an important new aspect of participative
democracy, reviewed in {\citep{Caba04a}. Our model casts this process as a
probabilistic voting problem, introduced first by \citet{Gibb77a} as a way
to design non dictatorial strategyproof decision rules. The literature he
inspired viewed randomization as a way around the defects of deterministic
rules, mostly to allow anonymous and neutral rules, or to circumvent the
absence of Condorcet winners~%
\citep[see e.g,
][]{Fish84a,LLL93b,ABBB15a,AzSt14a,Bran13a}. But recent work turns its
attention to mixtures of outcomes with time-sharing or compromise in mind:
see e.g., \citep{BMS05a,AzSt14a,Aziz13b,
Aziz17b, FGM16a, BNPS17a}.}

{Our works takes direct inspiration from \citet{BMS02a,BMS05a} who
introduced the model of randomised voting under dichotomous preferences. In
the same }mathematical {model we present several new results about new
normative requirements such as participation incentives, a decentralization
axiom, weaker forms of strategyproofness, and stronger forms of fairness.}

{Two of our rules, EGAL and NMP, maximize respectively a familiar social
welfare ordering and a classic collective utility function. The EGAL rule is
the lead mechanism in the related assignment model with dichotomous
preferences in \cite{BoMo04a}. In probabilistic voting, the Egalitarian
Simultaneous Reservation rule of \citet{AzSt14a} can be seen as an
adaptation of the Egalitarian rule.}

Recent literature emphasizes that the {NMP rule is central to the
competitive approach of the fair division of private commodities, whether
divisible or indivisible~\citep{CKM+16a,BMSY17a}. We find here a new
application of this rule in the public decision making context, closer in
spirit to Nash's original bargaining model~\citep{Nash50b}. Our results are
related to those of \cite{FGM16a}, who also propose the NMP rule for
budgetary participation, reinterpret this rule as a Lindahl equilibrium, and
discuss its computational complexity. They allow for more general
preferences than ours (in particular, full-fledged vNM utilities), and show
the Core Fair Share property (Corollary 1 Section 2.3) just like in
statement $i)$ of our Theorem 3. They do not discuss incentives properties
or any alternative rule. }

The rules CUT and RP are non welfarist, in that they do not maximize any
social welfare ordering. The RP rule is well known (and was discussed in 
\cite{BMS05a}), and CUT is a fairly simple twist on the random dictator
first introduced by \cite{Dudd15a} who noted that it is strategyproof but
did not develop its normative appeal.

{Fair Share is an early design constraint of decision mechanisms: see the
mathematical literature on cake cutting~\citep{Stei48a}, and on fair
division of microeconomic commodities~\citep{Moul03a,Vari74a,Thom15a}.}

{The group version of Fair Share captures the ubiquitous \textquotedblleft
protection of minorities\textquotedblright\ principle that is formally
related to cooperative stability in standard voting. It is also related to
the proportional veto principle \citep{Mou81a,Moul82a} and motivates
practical twists in the rules such as cumulative voting, especially
concerned with the protection of ethnic minorities in political elections~%
\citep{SaMa62a}, or minority stockholders in corporate governance %
\citep{Youn50a,Gord94a,PLSV00a}. See also the same concerns for EU
enlargement \cite{HuSa03b}. Our fairness notions are closely related to
proportional representation axioms in multi-winner voting as well~(see e.g., %
\citet{ABC+16a}).}

Strict {Participation has been considered in the deterministic voting model,
leading mostly to negative results. Our results complement those of
Brandl, Brandt, and Hofbauer (2015) who undertook a formal study of participation incentives in
probabilistic voting.}

\section{The Model}

A generic agent is $i\in N$, and $n=|N|$. A \textit{pure} public outcome is $%
a\in A$, and a \textit{mixture} of public outcomes is an element $z$ of the
simplex $\Delta (A)$, interpreted as a lottery over $A$, a profile of time
shares, or shares of other types of resources between the outcomes in $A$.
Both $N$ and $A$ are finite.

A utility function $u_{i}=(u_{ia})_{a\in A}$ is an element of $\{0,1\}^{A}$.
Agents who dislike all outcomes play no role in any or the rules we discuss,
thus we exclude them at once: the domain of preferences is $\Omega
=\{0,1\}^{A}\diagdown \{\mathbf{0\}}$, where $\mathbf{0}=0^{A}$; and $u\in
\Omega ^{N}$ is a profile of utility functions. In the examples we always
represent $u$ as a $N\times A$ matrix filled with $0$-s and $1$-s, and we
use the notation: $u_{S}=\sum_{i\in S}u_{i}$ and $u_{SB}=\sum_{i\in
S}\sum_{a\in B}u_{ia}$ for $S\subseteq N$ and $B\subseteq A$.

A \textit{problem} $M$ is a triple $M=(N,A,u)$ where $u\in \Omega ^{N}$.

Actual utilities (welfare) at $z$ are written $U_{i}=u_{i}\cdot z$, and the
corresponding utility profile is written $U=u\cdot z\in \lbrack 0,1]^{N}$.
The set of feasible utility profiles is $\Phi (M)=\{U=u\cdot z|z\in \Delta
(A)\}$. Given $U\in \Phi (M)$ we set $\varphi ^{-1}(U)=\{z\in \Delta
(A)|U=u\cdot z\}$.\smallskip

\textbf{Definition 1}

\noindent $i)$\textit{\ In problem }$M=(N,A,u)$\textit{\ a feasible utility
profile }$U\in \Phi (M)$\textit{\ is efficient\ if there is no profile }$%
U^{\prime }\in \Phi (M)$\textit{\ such that }$U\leq U^{\prime }$\textit{\
and at least one inequality }$U_{i}\leq U_{i}^{\prime }$\textit{\ is strict.}

\noindent $ii)$\textit{\ A mixture }$z\in \Delta (A)$\textit{\ is efficient
in }$M$\textit{\ if the profile }$u\cdot z$\textit{\ is efficient.}

\noindent $iii)$\textit{\ Fix }$\varepsilon \in \lbrack 0,1]$\textit{; the
profile }$U\in \Phi (M)$\textit{\ is }$\varepsilon $\textit{-efficient if
there exists }$U^{\prime }\in \Phi (M)$\textit{\ such that }$U\leq
\varepsilon U^{\prime }$\textit{.}\smallskip

\textbf{Definition 2}

\noindent $i)$ \textit{A rule }$F$\textit{\ picks one }$U\in \Phi (M)$%
\textit{\ for each problem }$M$\textit{; the mapping }$f$\textit{\ picks the
corresponding mixtures: }$f(M)=\varphi ^{-1}(F(M))$\textit{, so that }$%
F(M)=u\cdot f(M)$. \textit{Moreover }$F$\textit{\ and }$f$\textit{\ are
Anonymous (treat agents symmetrically) and Neutral (treat outcomes
symmetrically).}

\noindent $ii)$ \textit{The rule }$F$\textit{\ is efficient if it selects an
efficient profile in every problem. For any }$n$ \textit{the rule is }$%
\varepsilon (n)$\textit{-inefficient if }$a)$\textit{\ there exists a
problem }$M$\textit{\ of size }$n$ \textit{and a profile }$U\in \Phi (M)$%
\textit{\ such that }$F(M)$\textit{\ is }$\varepsilon (n)$\textit{%
-inefficient, and }$b)$\textit{\ no smaller number }$\varepsilon ^{\prime
}(n)$\textit{\ meets this property.}\smallskip

A rule is \textquotedblleft welfarist\textquotedblright\ by design, in the
sense that it does not distinguish between mixtures resulting in the same
utility profile. For instance if two outcomes $a,b$ are \textquotedblleft
clones\textquotedblright\ in problem $M$ (liked by exactly the same agents),
a rule is oblivious to shifting some weight from $a$ to $b$.

The efficient pure outcomes in $A$ are easy to recognize: $a$ is efficient 
\textit{if and only if} there is no $b$ such that the set of agents liking $%
b $ is strictly larger than the set liking $a$. We call such outcomes 
\textit{undominated.} In the following example

\begin{equation}
\begin{array}{cccccc}
N\downarrow A\rightarrow & a & b & c & d & e \\ 
1 & 0 & 0 & 0 & 1 & 1 \\ 
2 & 0 & 0 & 1 & 1 & 0 \\ 
3 & 1 & 1 & 0 & 0 & 0 \\ 
4 & 1 & 0 & 1 & 0 & 0 \\ 
5 & 0 & 1 & 0 & 1 & 1%
\end{array}
\label{3}
\end{equation}%
outcome $e$ is dominated by $d$, and the four other outcomes are
undominated. However, convex combinations of undominated outcomes may well
be inefficient. In the example, any mixture $z$ such that $z_{b},z_{c}$ are
both positive, say $z_{b},z_{c}\geq \alpha >0$, can be improved by
redistributing the weight $\alpha $ to $a$ and to $d$. That is, $z$ is
Pareto inferior to the mixture%
\[
z^{\prime }=(z_{a}+\alpha ,z_{b}-\alpha ,z_{c}-\alpha ,z_{d}+\alpha ,z_{e}). 
\]

Of special interest are those problems where any mixture of undominated pure
outcomes is efficient: in the probabilistic interpretation of our model this
means that ex post efficiency implies ex ante efficiency. Indeed the four
rules we discuss below mix only undominated outcomes, so in such problems
their efficiency is guaranteed.

In our first (minor) result, the set of outcomes liked by an agent is called
her \textit{like-set}.\smallskip

\textbf{Lemma }\textit{All mixtures of undominated (pure) outcomes are
efficient in problem }$M$ \textit{in two cases:}

\noindent $i)$ \textit{If }$|A|\leq 3$\textit{\ and/or }$|N|\leq 4$;

\noindent $ii)$ \textit{If }$A$ \textit{can be ordered in such a way that
the like-set of every agent is an interval.\smallskip }

Statement $i)$ is proven by \cite{Dudd15a}; it implies that example (\ref{3}%
) has the smallest sizes of $A$ and $N$ for which a combination of
undominated outcomes is inefficient.

\textbf{Proof of statement }$ii)$

\noindent Fix a problem $M$ as in statement $ii)$. If some outcomes are
\textquotedblleft clones\textquotedblright\ (liked by exactly the same set
of agents), a class of clones is an interval as well and it is clearly
enough to prove the statement for the \textquotedblleft
decloned\textquotedblright\ problem where each interval of clones has shrunk
to a single outcome. \ Thus we can assume that our problem has no clones.

Let $A^{\ast }$ denote the subset of undominated pure outcomes. We fix a
mixture $z$ with support in $A^{\ast }$ ($z\in \Delta (A^{\ast })$) and
assume some other mixture $y\in \Delta (A^{\ast })$ makes everyone weakly
better off than $z$: we will show $y=z$, which implies the statement.

We keep in mind that for any two $a,b$ in $A^{\ast }$ there is some agent $i$
who likes $a$ but not $b$, because $a$ and $b$ are not clones. Write the
ordered set $A^{\ast }$as $\{1,\cdots ,K\}$ and apply this remark to the
first two agents: some agent $i$ likes $1$ but not $2$, hence $i$ likes only 
$1$ and $u_{i}\cdot z\leq u_{i}\cdot y$ implies $z_{1}\leq y_{1}$. Some
agent $j$ likes $2$ but not $3$, hence $j$ likes $1,2$ or just $2$, so $%
u_{j}\cdot z\leq u_{j}\cdot y$ is either $z_{12}\leq y_{12}$ or $z_{2}\leq
y_{2}$ and either way we deduce $z_{12}\leq y_{12}$. Similarly there is some 
$k$ who likes $3$ but not $4$, so $u_{k}\cdot z\leq u_{k}\cdot y$ means that
at least one of $z_{3}$, $z_{23}$, and $z_{123}$ increases weakly and
inequality $z_{123}\leq y_{123}$ follows in each case. An obvious induction
argument gives%
\[
z_{12\cdots k}\leq y_{12\cdots k}\text{ for all }k,1\leq k\leq K 
\]%
The symmetric argument starting from outcome $K$ gives%
\[
z_{k(k+1)\cdots K}\leq y_{k(k+1)\cdots K}\text{ for all }k,1\leq k\leq K 
\]%
and the desired conclusion $y=z$ follows. $\blacksquare $

\section{Excludable Strategyproofness; the Egalitarian rule}

We start with the familiar prior-free incentive compatibility requirement
that misreporting one's preferences is never profitable if no agent can
coordinate this move with other agents.\footnote{%
Recall from Propositions 2 and 3 in \cite{BMS05a} that\ in our model group
versions of SP are not compatible with efficiency, even in the ex post sense.%
}

Notation: upon replacing in the profile $u$ the coordinate $u_{i}$ by
another $u_{i}^{\prime }\in \Omega $, the resulting profile is $%
(u|^{i}u_{i}^{\prime })$.%
\[
\text{\textbf{Strategyproofness }(SP)\textbf{:} }u_{i}\cdot f(M)\geq
\max_{z^{\prime }\in f(N,A,(u|^{i}u_{i}^{\prime }))}u_{i}\cdot z^{\prime }%
\text{ for all }M\text{, }i\text{ and }u_{i}^{\prime } . 
\]

The simplest strategyproof rule adapts approval voting to our model: it
selects only those outcomes liked by the largest number of agents. Write $%
\Phi ^{p}(M)$ for the set of utility profiles implemented by pure outcomes
in $A$: $\Phi ^{p}(M)=\{U\in \lbrack 0,1]|\exists a\in A\forall i\in
N,U_{i}=u_{ia}\}$. With the notation $avg(Y)$ for the uniform average
operation on a set $Y$ of utility profiles, we define the%
\[
\text{\textbf{Utilitarian rule (UTIL):}}\mathbf{\ }F^{ut}(M)=avg\{\arg
\max_{U\in \Phi ^{p}(M)}U_{N}\} . 
\]%
Note that the rule deliberately treats a problem with two identical columns
exactly as the reduced problem where only one column remains.

The careful reader can check that this defines a rule in the sense of
Definition 1, one that is efficient and strategyproof. However UTIL ignores
minority opinions entirely so it fails to address the normative concerns
described in the Introduction.

If an agent gets a fair $1/n$-th share of total decision power, she will use
it on an outcome she likes. We take the following lower bound on individual
welfare as the first test that mixing is fair:%
\[
\text{\textbf{Individual Fair Share }(IFS): }U=F(M)\Longrightarrow U_{i}\geq 
\frac{1}{n}\text{ for all }M\text{ and all }i . 
\]

The main result of \cite{BMS05a} is that a rule cannot be together
Efficient, Strategyproof, and meet the Individual Fair Share. Our first
result is that this impossibility disappears if we weaken SP as explained
below. To motivate this weakening, we adapt to our model the extremely
familiar idea of equalizing individual utilities while respecting efficiency.

The lexicographic ordering in $[0,1]^{\{1,\cdots ,n\}}$ maximizes the first
coordinate, and when this is not decisive, the second one, and so on. For a
utility profile $U\in \lbrack 0,1]^{N}$ the vector $U^{\ast }\in \lbrack
0,1]^{\{1,\cdots ,n\}}$ is obtained by rearranging its coordinates
increasingly. Then the leximin ordering $\succ _{leximin}$ compares $U^{1}$
and $U^{2}$ in $[0,1]^{N}$ exactly as the lexicographic ordering compares $%
U^{1\ast }$ and $U^{2\ast }$ in $[0,1]^{\{1,\cdots ,n\}}$.%
\[
\text{\textbf{Egalitarian rule (EGAL)}: }F^{eg}(M)=\arg \max_{U\in \Phi
(M)}\succ _{leximin} . 
\]%
This maximization yields a unique and efficient {utility profile} (see e.
g., Lemma 1.1 in \citep{Moul88a}). Anonymity and Neutrality are clear. To
check Individual fair Share, pick for each agent $i$ a pure outcome $a_{i}$
she likes, and observe that the uniform average of the $a_{i}$-s ensures
utility at least $1/n$ to each agent: therefore the egalitarian profile $%
U^{eg}$ must have $U_{1}^{eg\ast }\geq 1/n$.

Here is the simplest problem where the rule EGAL is vulnerable to a
misreport of preferences:%
\[
\text{true profile }u=%
\begin{array}{c|ccc}
N\downarrow A\rightarrow & a & b & c \\ 
1 & 1 & 1 & 0 \\ 
2 & 0 & 1 & 0 \\ 
3 & 0 & 0 & 1%
\end{array}%
\rightarrow \text{\ misreport }\widetilde{u}=%
\begin{array}{c|ccc}
N\downarrow A\rightarrow & a & b & c \\ 
1 & 1 & \widetilde{0} & 0 \\ 
2 & 0 & 1 & 0 \\ 
3 & 0 & 0 & 1%
\end{array}%
. 
\]%
At the true profile $u$ outcome $a$ is dominated and EGAL mixes $b$ and $c$, 
$z=(0,\frac{1}{2},\frac{1}{2})$. After the misreport by agent $1$, outcome $%
a $ no longer appears dominated and EGAL mixes equally the three outcomes, $%
\widetilde{z}=(\frac{1}{3},\frac{1}{3},\frac{1}{3})$. Agent $1$'s utility
raises from $1/2$ at $z$ to $2/3$ at $\widetilde{u}$, \textit{because he can
enjoy outcome }$b$\textit{\ despite pretending not to. }The latter is
avoidable if the public outcome are excludable: based on reported
preferences, the mechanism excludes agents from consuming outcomes they
claim to dislike. Recall the discussion of this possibility in the examples
of Section 1.

The following incentives property, where we use the notation $u_{i}\wedge
u_{i}^{\prime }$ for the coordinate-wise minimum of the two utility
functions, captures the resulting weaker incentive compatibility requirement:%
\[
\text{\textbf{Excludable Strategyproofness} (EXSP)} 
\]

\[
u_{i}\cdot f(M)\geq \max_{z^{\prime }\in f(N,A,(u|^{i}u_{i}^{\prime
}))}(u_{i}\wedge u_{i}^{\prime })\cdot z^{\prime }\text{ \ for all }M\text{, 
}i\text{ and }u_{i}^{\prime } . 
\]%
To make this definition more explicit, we identify the true utility $u_{i}$
by its \textit{like-set }$L_{i}=\{a\in A|u_{ia}=1\}$, and partition it as $%
L_{i}=L_{i}^{0}\cup L_{i}^{-}$. Agent $i$'s misreports is $L_{i}^{\prime
}=L_{i}^{0}\cup L_{i}^{+}$ where $L_{i}^{+}\subseteq A\diagdown L_{i}$: she
pretends to like $L_{i}^{+}$ and to dislike $L_{i}^{-}$. The like-set of $%
u_{i}\wedge u_{i}^{\prime }$ is $L_{i}^{0}$ therefore EXSP reads:%
\[
z_{L_{i}^{0}\cup L_{i}^{-}}\geq z_{L_{i}^{0}}^{\prime }\text{ \ for all }%
z\in f(M)\text{, }z^{\prime }\in f(N,A,(u|^{i}u_{i}^{\prime })) . 
\]%
It is useful to decompose EXSP in two statements. In the first one agent $i$
misreports only by inflating her like-set ($L_{i}=L_{i}^{0}$):%
\[
\text{\textbf{SP}}^{+}\text{: }u_{i}\cdot f(M)\geq \max_{z^{\prime }\in
f(N,A,(u|^{i}u_{i}^{\prime }))}u_{i}\cdot z^{\prime }\text{ \ for all }M%
\text{, }i\text{ and }u_{i}^{\prime }\text{ s. t. }u_{i}\leq u_{i}^{\prime }
. 
\]%
and in the second one, only by decreasing this set ($L_{i}^{+}=\varnothing $%
):%
\[
\text{\textbf{SP}}^{-}\text{: }u_{i}\cdot f(M)\geq \max_{z^{\prime }\in
f(N,A,(u|^{i}u_{i}^{\prime }))}u_{i}^{\prime }\cdot z^{\prime }\text{ \ for
all }M\text{, }i\text{ and }u_{i}^{\prime }\text{ s. t. }u_{i}^{\prime }\leq
u_{i} . 
\]

That EXSP equals the combination of SP$^{+}$ and SP$^{-}$ is clear by
applying first SP$^{-}$ from $u_{i}$ to $u_{i}\wedge u_{i}^{\prime }$, then
SP$^{+}$ from $u_{i}\wedge u_{i}^{\prime }$ to $u_{i}^{\prime }$. Similarly
SP is the combination of SP$^{+}$ and SP$^{\ast }$:%
\[
\text{\textbf{SP}}^{\ast }\text{: }u_{i}\cdot f(M)\geq \max_{z^{\prime }\in
f(N,A,(u|^{i}u_{i}^{\prime }))}u_{i}\cdot z^{\prime }\text{ \ for all }M%
\text{, }i\text{ and }u_{i}^{\prime }\text{ s. t. }u_{i}^{\prime }\leq u_{i}
. 
\]%
and the above example shows that EGAL violates SP$^{\ast }$.\smallskip

\textbf{Theorem} \textbf{1 }\textit{The Egalitarian rule is Efficient,
Excludable Strategyproof, and guarantees Individual Fair Shares.\smallskip }

\textbf{Proof}

\noindent \textit{Preliminary notation and remarks}: If $M\subseteq N$ and $%
U\in \lbrack 0,1]^{M}$ then $U^{\ast }$ is the set of \textit{distinct}
coordinates $U^{\ast k}$ of $U$ arranged increasingly; so $U^{\ast }$ may be
of lower dimension than $M$.

Fix a problem $M=(N,A,u)$. For any $M\subseteq N$ and convex compact $%
C\subseteq \Delta (A)$ the projection on $M$ of the set of feasible utility
profiles $\Phi (C)=\{U=u\cdot z|z\in C\}$ is convex and compact, so it
admits a unique leximin optimal element that we write $F^{eg}(M,C,u)\in
\lbrack 0,1]^{M}$. This extends the domain of the mapping $F^{eg}$, and note
that we abuse notation by keeping $u$ instead of its restriction to $M\times
A$.

Recall the algorithm defining $U=F^{eg}(M,C,u)$. Start with 
\[
U^{\ast 1}=\max_{z\in C}\min_{j\in M}\{u_{j}\cdot z\}. . 
\]%
Write $N^{1}$ for the set of agents achieving this minimum, $%
P^{1}=N\diagdown N^{1}$, and $C^{1}=\{z\in C|u_{j}\cdot z=U^{\ast 1}$ for
all $j\in N^{1}\}$. We stop if $N^{1}=N$, otherwise we set $U^{\ast
2}=\max_{z\in C^{1}}\min_{j\in P^{1}}\{u_{j}\cdot z\}$. We let $N^{2}$ be
the set of agents achieving $U^{\ast 2}$, $P^{2}=N\diagdown (N^{1}\cup
N^{2}) $, and $C^{2}$ the subset of $C^{1}$ achieving $U^{\ast 2}$ in $N^{2}$%
; we stop if $P^{2}=\varnothing $, otherwise we set $U^{\ast 3}=\max_{z\in
C^{2}}\min_{j\in P^{2}}\{u_{j}\cdot z\}$, and so on. We end up with a
partition $N={\Large \cup }_{k=1}^{K}N^{k}$ such that $U_{i}$ equals $%
U^{\ast k}$ whenever $i\in N^{k}$.

Turning to the proof of statement $i)$, we saw that it is enough to show
separately SP$^{-}$ and SP$^{+}$. Fix an arbitrary $M=(N,A,u)$. An agent who
likes all outcomes, $u_{i}=\mathbf{1}$, cannot benefit from any misreport;
pick now $i\in N$ such that $u_{ia}=0$ for at least one $a$, and a profile $%
\widetilde{u}$ identical to $u$ for all $j\in N\diagdown i$ and such that $%
u_{i}\lneqq \widetilde{u}_{i}$ (so at least one $0$ in $u_{i}$ is changed to
a $1$). Let $U=F^{eg}(N,A,u)$ and $\widetilde{U}=F^{eg}(N,A,\widetilde{u})$
be implemented respectively by some lotteries $z$ and $\widetilde{z}$. We
prove successively:%
\begin{equation}
\widetilde{U}_{i}\geq U_{i}  \label{9}
\end{equation}%
\begin{equation}
U_{i}=u_{i}\cdot z\geq u_{i}\cdot \widetilde{z}  \label{10}
\end{equation}%
The first inequality implies SP$^{-}$ (when $i$ with true $\widetilde{U}_{i}$
reports $U_{i}$), the second gives SP$^{+}$ (when $i$ with true $U_{i}$
reports $\widetilde{U}_{i}$).

We clearly have $\widetilde{U}\succeq _{leximin}U$, in particular $%
\widetilde{U}^{\ast 1}\geq U^{\ast 1}$: this proves (\ref{9}) if $%
U_{i}=U^{\ast 1}$. Assume for the rest of the proof $U_{i}=U^{\ast \ell }$
where $\ell \geq 2$. We check first $\widetilde{U}^{\ast 1}=U^{\ast 1}$. If $%
\widetilde{U}^{\ast 1}>U^{\ast 1}$ we pick $\varepsilon \in ]0,1]$, and note
that the mixture $z^{\prime }=\varepsilon \widetilde{z}+(1-\varepsilon )z$
ensures $u_{j}\cdot z^{\prime }>U^{\ast 1}$ for all $j\in N\diagdown i$; for 
$\varepsilon $ small enough we also have $u_{i}\cdot z^{\prime }>U^{\ast 1}$
because $u_{i}\cdot z>U^{\ast 1}$. This contradicts the definition of $%
U^{\ast 1}$.

Set $N^{1}=\{j|U_{j}=U^{\ast 1}\}$ and $\widetilde{N}^{1}=\{j|\widetilde{U}%
_{j}=U^{\ast 1}\}$. We use a similar argument to show next $N^{1}\subseteq 
\widetilde{N}^{1}$. If $j\in N^{1}$ and $u_{j}\cdot \widetilde{z}>U^{\ast 1}$%
, then for any $\varepsilon \in ]0,1]$ the mixture $z^{\prime }=\varepsilon 
\widetilde{z}+(1-\varepsilon )z$ gives $u_{k}\cdot z^{\prime }\geq U^{\ast
1} $ for all $k\in N\diagdown \{j,i\}$ and $u_{j}\cdot z^{\prime }>U^{\ast
1} $; for $\varepsilon $ small enough we also have $u_{i}\cdot z^{\prime
}>U^{\ast 1}$ (because $U_{i}=U^{\ast \ell }>U^{\ast 1}$) and then $%
z^{\prime }$ guarantees exactly $U^{\ast 1}$ to a smaller set of agents than 
$z$, and strictly more to all others. This implies that $u\cdot z^{\prime }$
leximin-dominates $u\cdot z$, contradiction.

Similarly the strict inclusion $N^{1}\varsubsetneq \widetilde{N}^{1}$ would
imply that the vector $\widetilde{U}$ is strictly leximin-dominated by $U$,
which we saw is not true.

So far we have shown that the maxi-minimization of feasible utilities -- the
first step in the algorithm defining the leximin solution-- gives at $u$ and 
$\widetilde{u}$ identical values $U^{\ast 1}$ and $\widetilde{U}^{\ast 1}$,
and identical sets $N^{1}$ and $\widetilde{N}^{1}$. Now the second step of
the algorithms, delivering $U^{\ast 2}$, $\widetilde{U}^{\ast 2}$, and $%
N^{2} $, $\widetilde{N}^{2}$, is the same maxi-minimization problem applied
in both cases to $C^{1}=\{z\in \Delta (A)|u_{j}\cdot z=U^{\ast 1}$ for all $%
j\in N^{1}\}$ and $P^{1}=N\diagdown N^{1}$. Mimicking the above proof we
deduce that, if $U_{i}=U^{\ast 2}$ then $\widetilde{U}_{i}\geq U_{i}$, and
if $U_{i}=U^{\ast \ell }$ for some $\ell \geq 3$, then $U^{\ast 2}=%
\widetilde{U}^{\ast 2}$, $N^{2}=\widetilde{N}^{2}$. The induction argument
establishing $U^{\ast k}=\widetilde{U}^{\ast k}$, $N^{k}=\widetilde{N}^{k}$
up to $k=\ell -1$, and finally (\ref{9}) is now clear.\medskip

To prove (\ref{10}) we compare the profiles $u\cdot z$ and $u\cdot 
\widetilde{z}$. We just saw that they coincide on $N\diagdown P^{\ell -1}=%
{\Large \cup }_{k=1}^{\ell -1}N^{k}$, and that if a mixture guarantees
utility $U^{\ast k}$ to all agents in $N^{k}$ for $k=1,\cdots ,\ell -1$, it
cannot guarantee (at $u$) more than $U^{\ast \ell }$ to all agents in $%
P^{\ell -1}$: $z$ and $\widetilde{z}$ are two such lotteries, so if $%
u_{i}\cdot \widetilde{z}>u_{i}\cdot z=U^{\ast \ell }$, there is some $j\in
P^{\ell -1}$ for whom $u_{j}\cdot \widetilde{z}<U^{\ast \ell }$. But $%
\widetilde{U}^{\ast \ell }\geq U^{\ast \ell }$ (because $\widetilde{U}$
weakly leximin-dominates $U$) and $\widetilde{u}_{j}\cdot \widetilde{z}%
=u_{j}\cdot \widetilde{z}\geq \widetilde{U}^{\ast \ell }$, thus we reach a
contradiction. $\blacksquare $

\section{Strict Participation and Unanimous Fair Share}

A striking feature of the Egalitarian rule is \textit{Clone Invariance}: if
at least one voter who shares my preferences does vote, adding my own vote
will not change the resulting mixture. This holds because, fixing an agent $%
i $, the leximin ordering compares two utility profiles $U$ and $U^{\prime }$
in the same way as $\widetilde{U}$ and $\widetilde{U^{\prime }}$, where from 
$V$ to $\widetilde{V}$ we add an $(n+1)$-th coordinate repeating $V_{i}$.
Thus the rule is oblivious to the size of support for a particular
preference, an unpalatable feature in all the examples discussed in the
introduction.

We now define two requirements capturing, each in a different way, the
concern that numbers should matter. The first one is an incentive property.

Given a problem $M$ and agent $i$, define $M(-i)=(N\diagdown i,A,u_{-i})$
and $U_{i}(-i)=\max_{z\in f(M(-i))}u_{i}\cdot z$. 
\[
\text{\textbf{Participation }(PART)\textbf{:} }F_{i}(M)\geq U_{i}(-i)\text{
for all }M\text{ and }i . 
\]%
The violation of Participation is commonly called the No Show Paradox~{%
\citep{BrFi83a}}: a voter is better off abstaining to go to the polls. In
the context of budgetary participation, we want more: everyone should have a
strict incentive to show up, lest many agents, feeling disenfranchised, will
stay home or put a blank ballot, and the result of the vote will not give an
accurate picture of the opinion profile.%

\[
\text{ \textbf{Strict Participation }(PART}^{\ast }\text{):} 
\]%
\[
F_{i}(M)\geq U_{i}(-i)\text{ and }\{U_{i}(-i)<1\Longrightarrow
F_{i}(M)>U_{i}(-i)\}\text{ for all }M\text{ and }i . 
\]%

Under dichotomous preferences that we consider, strong SD-participation and SD-participation as studied by \citet{BBH15b} coincide with PART and very strong SD-participation coincides with PART$^{\ast }$. 
A consequence of PART$^{\ast }$ is \textit{Clone Responsiveness}: I am
strictly better off if one or more agents with preferences identical to mine
cast their vote. Thus the Egalitarian rule violates PART$^{\ast }$, although
it satisfies PART.\footnote{%
Define $U^{\ast }=\arg \max_{U\in \Phi (M)}\succ _{leximin}$ ; $\overline{U}%
_{-1}=\arg \max_{U\in \Phi (M(-1))}\succ _{leximin}$ and $\overline{U}%
_{1}=U_{1}(-1)$. If $\overline{U}_{1}>U_{1}^{\ast }$ we have successively $%
\overline{U}\succeq _{leximin}(\overline{U}_{1},U^{\ast })$ then $(\overline{%
U}_{1},U^{\ast })\succ _{leximin}U^{\ast }$, contradiction.}

The second axiom, in the spirit of cumulative voting, allows groups of
agents with identical preferences to pool their respective shares of
decision power. This leads to the following strengthening of IFS, where we
set again $U=F(M)$:%
\[
\text{\textbf{Unanimity Fair Share }(UFS)}: 
\]%
\[
\text{for all }S\subseteq N\text{: }\{u_{i}=u_{j}\text{ for all }i,j\in
S\}\Longrightarrow U_{i}\geq \frac{|S|}{n}\text{ for all }i\in S . 
\]

In the statement of UFS the unanimous group $S$ can be a minority or a
majority. However unanimous preferences are much more likely in small than
large groups, so this property will be more relevant in practice to
minorities.

All three rules discussed in the next two sections meet Strict Participation
and Unanimous Fair Share. Thus they cannot be both efficient and
strategyproof. We start with two strategyproof rules.

\section{Incentive compatibility and Fairness; the Conditional Utilitarian
rule}

We introduce two rules adapting to our model the familiar random dictator
mechanism (see \cite{Gibb77a}). The difficulty is the treatment of
indifferences: if I can dictate the outcome for a $1/n$-th share of the
time, how should I choose in my like-set ?

The first rule, introduced by \cite{Dudd15a}, applies a simple utilitarian
test: I focus on the outcomes liked by the largest number of other agents.
Consider the set $\Phi ^{p}(M;i)=\{U\in \Phi ^{p}(M)|U_{i}=1\}$ of all the
utility profiles corresponding to the like-set of agent $i$. Each agent
spreads her share $\frac{1}{n}$ equally between the profiles in $\Phi
^{p}(M;i)$ with maximal support:%
\[
\text{\textbf{Conditional Utilitarian} (CUT) rule: }F^{cut}(M)=\frac{1}{n}%
\sum_{i\in N}avg\{U|U\in \arg \max_{U^{\prime }\in \Phi
^{p}(M;i)}U_{N}^{\prime }.\} 
\]

\textit{Remark 1. Our definition of the domain }$\Omega $\textit{\ allows
for agents who like all outcomes, }$u_{i}=1^{A}$\textit{. The presence of
such agents is of no consequence for the rules UTIL, EGAL, RP and NMP, but
it does impact the mixture selected by the CUT rule, as such agents put
their weight on the utilitarian outcomes (those with largest support).
Suppose we choose to exclude those agents in the definition of the CUT rule:
this will not affect the incentives and fairness properties of the rule
identified below, nor its Decentralization property in Section 8.\smallskip }

The next rule uses a familiar hierarchical rule to resolve indifferences,
that plays a critical role in probabilistic voting (\citep{ABB13b}), as well
as for assigning indivisible private goods (\citep{AbSo98a}, \citep{BoMo01a}%
). Let $\Theta (N)$ be the set of strict orderings $\sigma $ of $N$. For any 
$\sigma \in \Theta (N)$ the $\sigma $-\textit{Priority rule} $F^{\sigma }$
guarantees full utility to agent $\sigma (1)$; next to agent $\sigma (2)$ as
well if $1$ and $2$ like a common outcome, else $\sigma (2)$ is deemed
irrelevant; next to agent $\sigma (3)$ if she likes an outcome that all
relevant agents before her like, else she is irrelevant; and so on.%
\[
\text{\textbf{Random Priority rule (RP)}} 
\]%
\[
F^{rp}(M)=\frac{1}{n!}\sum_{\sigma \in \Theta (N)}F^{\sigma }(M)\text{ where 
}F^{\sigma }(M)=\arg \max_{U\in \Phi (M)}\succ _{lexico}^{\sigma } . 
\]

If mixtures in $\Delta (A)$ represent lotteries, the RP rule picks an
ordering $\sigma $ with uniform probability and computes $U^{\sigma }$. But
in other interpretations, time shares or the distribution of other
resources, this simple implementation is not available. We retain
nevertheless the intuitive probabilistic terminology.

After checking that both rules are incentive compatible and fair, we compare
them from the efficiency angle, and recap our discussion in Theorem 2 below.

Clearly each $\sigma $-Priority rule $F^{\sigma }$ is strategyproof, and SP
is preserved by convex combinations, thus RP is strategyproof as well.
Checking PART$^{\ast }$ is equally easy: to each ordering $\widetilde{\sigma 
}$ of $N\diagdown i$ we associate the $n$ orderings $\sigma $ of $N$ where $%
i $ can have any rank from first to last: agent $i$ is weakly better off at $%
F^{\sigma }(M)$ than at $F^{\widetilde{\sigma }}(M(-i))$, strictly so if he
is gets utility $0$ in $F^{\widetilde{\sigma }}(M(-i))$; thus the only case
where $i$ does not strictly benefit by showing up is when he gets utility $1$
in $F^{rp}(M(-i))$. For UFS, it is enough to observe that a member of
coalition $S$ is first in $\sigma $ with probability $\frac{|S|}{n}$.

Check now that CUT meets the same three properties. UFS is clear. We
decompose SP into the combination of SP$^{+}$ and SP$^{\ast }$. Start with SP%
$^{\ast }$: with the notations just before Theorem 1, assume agent $i$'s
like-set is $L_{i}=L_{i}^{0}\cup L_{i}^{-}$, and he reports $L_{i}^{0}$
instead. Fix another agent $j$ and check that the potential change in the
way $j$ uses her $1/n$-th share does not hurt $i$. If $j$ was putting no
weight in $L_{i}^{-}$, then she still loads exactly the same set. If $j$ was
only loading (a subset of) $L_{i}^{-}$, she was helping $i$ as much as
possible and cannot do more after the change. If $j$ was loading some
outcomes in $L_{i}^{-}$, some in $B\subseteq L_{i}^{0}$ and some in $%
C\subseteq A\diagdown L_{i}$, where $B\cup C\neq \varnothing $, then she
redistributes all her weight on $L_{i}^{-}$ uniformly in $B\cup C$ and this
clearly cannot benefit agent $i$. Turning to SP$^{-}$ we assume $i$ likes $%
L_{i}$ and reports $L_{i}^{\prime }=L_{i}\cup L_{i}^{+}$ instead. If $j$ was
putting some load in $L_{i}^{+}$, she now loads only $L_{i}^{+}$, so she
does not help $i$ at all. If she was not loading $L_{i}^{+}$ at all, she may
do so now, in which case some weight will be taken uniformly from the set
she was loading before: this cannot benefit $i$ strictly.

Check PART$^{\ast }$. Fix a problem $M$, an agent $i$, and for every $j\in
N\diagdown i$ let $B_{j}$ be the set of outcomes agent $j$ loads in problem $%
M(-i)$. Set $N^{+}=\{j\in N\diagdown i|B_{j}\cap L_{i}\neq \varnothing \}$
and $N^{-}=N\diagdown (N^{+}\cup i)$. Before participating agent $i$'s
utility was%
\[
\frac{1}{n-1}\sum_{j\in N^{+}}\lambda _{j}\text{ where }\lambda _{j}=\frac{%
|B_{j}\cap L_{i}|}{|B_{j}|} . 
\]%
After $i$ shows up every $j$ in $N^{+}$ loads only $B_{j}\cap L_{i}$, and
agents in $N^{-}$ may give some of their load to $L_{i}$ therefore $i$'s
utility is at least $\frac{1}{n}(1+|N^{+}|)$. The inequality%
\[
\frac{1}{n-1}\sum_{j\in N^{+}}\lambda _{j}\leq \frac{|N^{+}|}{n-1}\leq \frac{%
1}{n}(1+|N^{+}|) . 
\]%
proves PART. And both inequalities are equalities if and only if each $%
\lambda _{j}=1$ and $|N^{+}|=n-1\Leftrightarrow N^{+}=N\diagdown i$; the
latter implies that $i$'s utility is already $1$ in $M(-i)$.

Example (\ref{3}) above shows that both RP and CUT are inefficient. Under
the CUT rule agents $1,2$ and $5$ load only $d$, while agent $3$ spreads his
load between $a$ and $b$, and agent $4$ between $a$ and $c$, resulting in
the mixture $z^{cut}=(\frac{1}{5},\frac{1}{10},\frac{1}{10},\frac{3}{5},0)$.
Under RP we get $z^{rp}=(\frac{1}{5},\frac{1}{6},\frac{1}{6},\frac{7}{15},0)$%
; for instance $b$ is selected in two cases only: if $3$ is first, and $5$
comes before $4$ (proba. $\frac{1}{10}$), or $5$ is first and $3$ is first
among $1,2,3$ (proba. $\frac{1}{15}$). As noted at the end of Section 3,
shifting the weight of $b$ and $c$ to $a$ and $d$ is a Pareto improvement.
Clearly, then, $z^{rp}$ is more inefficient than $z^{cut}$.

In our next example, with $n=6$ and $|A|=5$,

\begin{equation}
\begin{array}{cccccc}
N\downarrow A\rightarrow & a & b & c & d & e \\ 
1 & 1 & 0 & 0 & 1 & 0 \\ 
2 & 1 & 0 & 0 & 0 & 1 \\ 
3 & 0 & 1 & 0 & 1 & 0 \\ 
4 & 0 & 1 & 0 & 0 & 1 \\ 
5 & 0 & 0 & 1 & 1 & 0 \\ 
6 & 0 & 0 & 1 & 0 & 1%
\end{array}
\label{5}
\end{equation}%
the CUT rule selects the efficient mixture $z^{cut}=(0,0,0,\frac{1}{2},\frac{%
1}{2})$ and $U_{i}^{cut}=0.5$ for all $i$, while RP picks $z^{rp}=(\frac{1}{9%
},\frac{1}{9},\frac{1}{9},\frac{1}{3},\frac{1}{3})$ and $U_{i}^{rp}=0.44$
for all $i$: thus $z^{cut}$ is strictly Pareto superior to $z^{rp}$.

The reverse situation cannot happen: the RP mixture never Pareto dominates
the CUT one. This follows because in all problems, total utility under RP is
at most that under CUT: $U_{N}^{rp}\leq U_{N}^{cut}$. Indeed $U^{cut}$ is
the uniform average of profiles $U(i)$ maximizing total utility in $\Phi
^{p}(M;i)$, and for each ordering $\sigma $ where $i$ is first, the
corresponding profile $U^{\sigma }$ is in $\Phi ^{p}(M;i)$ as well.\footnote{%
A consequence of this remark is that CUT and RP pick the same utility
profile at problem $M$ if and only if all undominated outcomes of $M$ are
liked by the same number of agents.}

We prove finally that \textit{the CUT rule is efficient more often than RP}%
:\ whenever RP picks an efficient mixture, so does CUT. Observe first that
both rules only give weight to undominated pure outcomes. In the case \ of
RP every such outcome $a$ has a positive weight, because it is selected
whenever the set of agents who like $a$ has the highest priority. Thus the
support of the RP mixture is exactly the set of all undominated columns.
Therefore RP selects an efficient mixture if and only if all mixtures with
support in this set are efficient as well\textbf{. }The claim follows
because the CUT rule is also a combination of undominated columns.

The next result adds to the discussion above some worst case computations
reinforcing the strong efficiency advantage of CUT over RP.\smallskip

\textbf{Theorem 2}

\noindent $i)$ \textit{Both rules CUT and RP are strategyproof and meet
Strict Participation and Unanimity Fair Share.}

\noindent $ii)$ \textit{Total utility at the CUT mixture is never below that
at the RP mixture, and the former may Pareto dominate the latter. If RP
picks an efficient mixture at some problem }$M$\textit{, so does CUT.}

\noindent $iii)$ \textit{The CUT rule is }$\varepsilon ^{cut}(n)$\textit{%
-efficient with }$\varepsilon ^{cut}(n)=O(n^{-\frac{1}{3}})$ and for all $%
n\geq 5$ we have%
\begin{equation}
\varepsilon ^{cut}(n)\geq \frac{1}{n}+(1-\frac{1}{n^{\frac{1}{3}}})\frac{3}{%
n^{\frac{1}{3}}}  \label{13}
\end{equation}%
\textit{The RP rule is }$\varepsilon ^{rp}(n)$\textit{-efficient with }$%
\varepsilon ^{rp}(n)\leq O(\frac{\ln (n)}{n})$.

\noindent $iv)$ \textit{The CUT mixture is computed in time polynomial in }$%
n+|A|$\textit{; computing the RP rule is \#P-complete in }$n+|A|$\textit{%
.\smallskip }

Recall from the Lemma in Section 3 that both CUT and RP are efficient if $%
n\leq 4$. For small values of $n$, the lower bound (\ref{13}) implies a high
guaranteed efficiency of CUT, a lower bound on $\varepsilon ^{cut}(n)$, and
the computations in Step 2 of the proof below yield a much smaller worst
case efficiency of RP, an upper bound on $\varepsilon ^{rp}(n)$:%
\[
\begin{array}{cccccccc}
n & 6 & 8 & 12 & 32 & 64 & 1024 & 16384 \\ 
\varepsilon ^{cut}\geq  & 91\% & 87\% & 82\% & 68\% & 58\% & 27\% & 11\% \\ 
\varepsilon ^{rp}\leq  & 83\% & 72\% & 64\% & 40\% & 24\% & 3\% & 0.12\%%
\end{array}%
\]

Together statements $ii)$ to $iv)$ make a very strong case that in our model
the CUT rule is a much more efficient interpretation of the random dictator
idea than RP.\smallskip

\textbf{Proof of statement }$iii)$ (Statement $iv)$ is explained in Section
9)

\noindent \textit{Step 1 Worst case inefficiency of CUT}

\textit{Step 1.a: }We construct a problem with large $n$ where the CUT
profile is $O(n^{-\frac{1}{3}})$ inefficient. We fix $N$ of size $n$, a
partition $N=N_{1}\cup N_{2}$, and an integer $p$ such that%
\[
p<n_{1},n_{2}\text{ and }n_{1}\text{ divides }(p-1)n_{2}\text{ where }%
n_{i}=|N_{i}|,i=1,2 . 
\]%
Problem $M$ has $2n_{2}+1$ outcomes labeled as $A=\{a\}\cup B\cup C$, where $%
B=\{b_{j},j\in N_{2}\}$ and $C=\{c_{j},j\in N_{2}\}$. Setting $%
(p-1)n_{2}=qn_{1}$, each agent $i\in N_{1}$ likes $a$, exactly $q$ outcomes
in $B$, and none in $C$; and each $j\in N_{2}$ dislikes $a$, likes only
outcome $b_{j}$ in $B$, and exactly $p-1$ outcomes in $C$. Moreover the
problem is symmetric in $N_{1}$ and in $N_{2}$, which can be achieved by
arranging cyclically the like-sets of the $N_{1}$ agents in $B$ and the
like-sets of the $N_{2}$ agents in $C$. Here is an example with $%
n_{1}=n_{2}=5,p=4$ and $q=3$, and the top five agents form $N_{1}$:%
\[
\begin{array}{ccccccccccc}
a & b_{1} & b_{2} & b_{3} & b_{4} & b_{5} & c_{1} & c_{2} & c_{3} & c_{4} & 
c_{5} \\ 
1 & 1 & 0 & 0 & 1 & 1 & 0 & 0 & 0 & 0 & 0 \\ 
1 & 1 & 1 & 0 & 0 & 1 & 0 & 0 & 0 & 0 & 0 \\ 
1 & 1 & 1 & 1 & 0 & 0 & 0 & 0 & 0 & 0 & 0 \\ 
1 & 0 & 1 & 1 & 1 & 0 & 0 & 0 & 0 & 0 & 0 \\ 
1 & 0 & 0 & 1 & 1 & 1 & 0 & 0 & 0 & 0 & 0 \\ 
0 & 1 & 0 & 0 & 0 & 0 & 1 & 0 & 0 & 1 & 1 \\ 
0 & 0 & 1 & 0 & 0 & 0 & 1 & 1 & 0 & 0 & 1 \\ 
0 & 0 & 0 & 1 & 0 & 0 & 1 & 1 & 1 & 0 & 0 \\ 
0 & 0 & 0 & 0 & 1 & 0 & 0 & 1 & 1 & 1 & 0 \\ 
0 & 0 & 0 & 0 & 0 & 1 & 0 & 0 & 1 & 1 & 1%
\end{array}%
\]%
Note that each outcome $b_{j}$ is liked by exactly $p$ agents, all but one
of them in $N_{1}$, and each $c_{j}$ is liked by exactly $p-1$ agents, all
in $N_{2}$.

Under the CUT rule, each agent $i\in N_{1}$ loads only $a$ because $n_{1}>p$%
, so $z_{a}=\frac{n_{1}}{n}$, and each $j\in N_{2}$ loads only $b_{j}$ so $%
z_{b_{j}}=\frac{1}{n}$; there is no weight on $C$. Total utility in each
group is%
\[
U_{N_{1}}=\frac{(n_{1})^{2}}{n}+\frac{n_{2}}{n}(p-1)\text{ ; }U_{N_{2}}=%
\frac{n_{2}}{n} 
\]%
and by the symmetries these are equally shared in $N_{1}$ and $N_{2}$
respectively.

Now consider the mixture $z^{\prime }$: $z_{a}^{\prime }=\frac{2}{3}$, $%
z_{c_{j}}^{\prime }=\frac{1}{3n_{2}}$ for all $j\in N_{2}$, and zero weight
on $B$, resulting in the total utilities%
\[
U_{N_{1}}^{\prime }=\frac{2}{3}n_{1}\text{ ; }U_{N_{2}}^{\prime }=\frac{1}{3}%
(p-1) 
\]%
again equally shared in each $N_{i}$.

For $n$ large enough we can pick $n_{1}$ and $p$ such that $n_{1}\simeq n^{%
\frac{2}{3}}$ and $p-1\simeq n^{\frac{1}{3}}$ (if $n$ is a cube these values
are exact and $q=n^{\frac{2}{3}}-n^{\frac{1}{3}}$) so that $\frac{n_{2}}{n}%
\simeq 1$. This yields the ratios%
\[
\frac{U_{N_{1}}^{\prime }}{U_{N_{1}}}\simeq \frac{\frac{2}{3}n^{\frac{2}{3}}%
}{2n^{\frac{1}{3}}}=\frac{1}{3}n^{\frac{1}{3}}=\frac{U_{N_{2}}^{\prime }}{%
U_{N_{2}}} 
\]%
and completes the proof of Step 1.a.\smallskip

\textit{Step 1.b. For an arbitrary problem }$M$ \textit{we give an
upper-bound of the inefficiency of the CUT mixture.}

We fix a problem $M$ and partition the agents according to their scores $%
\max_{U\in \Phi ^{p}(M;i)}U_{N}$, i.e., the utilitarian score of the
outcomes on which they spread their weight under the CUT rule. Let $%
p_{1}>p_{2}>\cdots >p_{K}>0$ be the sequence of such scores and $N_{k}$ the
subset of agents who load outcomes with score $p_{k}$. Note that $n_{1}\geq
p_{1}$. Set $A_{k}$ to be the set of outcomes loaded by at least one agent
in $N_{k}$: they all have the same score $p_{k}$ so the $A_{k}$-s are
pairwise disjoint. Note also that agents in $N_{k}$ do not like any outcome
in $A_{\ell }$ for $\ell <k$.

Consider finally the outcomes $b$ in $B=A\diagdown {\Large (\cup }%
_{1}^{K}A_{k}{\Large )}$, if any. Their utilitarian score $u_{N_{b}}$ is at
most $p_{1}-1$. We partition $B$ by gathering in $B_{k}$ all the outcomes
with a score in $[p_{k+1},p_{k}]$, with the convention $p_{K+1}=0$.
Therefore the agents in $N_{k}$ do not like any outcome in $B_{\ell }$ for $%
\ell <k$.

We prove first that for any feasible profile $U\in \Phi (M)$, we can find
convex weights $\pi _{1},\cdots ,\pi _{K}$ such that%
\begin{equation}
U_{N_{k}}\leq \pi _{k}p_{k}\text{ for }k=1,\cdots ,K  \label{11}
\end{equation}%
Pick $z\in \Delta (A)$ implementing $U$ and write for simplicity $%
z_{A_{k}}=x_{k}$ and $z_{B_{k}}=y_{k}$. The total contribution\footnote{%
Recall our notation $u_{SB}=\sum_{i\in S}\sum_{a\in B}u_{ia}$.} $%
U_{NA_{k}}=x_{k}p_{k}$ of $A_{k}$ to $U_{N}$ is shared between the agents of 
${\Large \cup }_{1}^{k}N_{\ell }$ only, so there are some convex weights $%
\gamma _{\ell }^{k},1\leq \ell \leq k,$ such that%
\[
U_{N_{\ell }A_{k}}=\gamma _{\ell }^{k}x_{k}p_{k}\text{ for all }1\leq \ell
\leq k\leq K . 
\]%
Similarly the contribution $U_{NB_{k}}$ of $B_{k}$ is shared in ${\Large %
\cup }_{1}^{k}N_{\ell }$ and $U_{NB_{k}}\leq y_{k}p_{k}$. So we can find
convex weights $\delta _{\ell }^{k},1\leq \ell \leq k,$ such that%
\[
U_{N_{\ell }B_{k}}\leq \delta _{\ell }^{k}y_{k}p_{k}\text{ for all }1\leq
\ell \leq k\leq K . 
\]%
Combining the above equality and inequality we have for all $k$%
\[
U_{N_{k}}=\sum_{\ell =k}^{K}(U_{N_{k}A_{\ell }}+U_{N_{k}B_{\ell }})\leq
\sum_{\ell =k}^{K}(\gamma _{k}^{\ell }x_{\ell }+\delta _{k}^{\ell }y_{\ell
})p_{\ell }\leq p_{k}\sum_{\ell =k}^{K}(\gamma _{k}^{\ell }x_{\ell }+\delta
_{k}^{\ell }y_{\ell }) 
\]%
so the weights $\pi _{k}=\sum_{\ell =k}^{K}(\gamma _{k}^{\ell }x_{\ell
}+\delta _{k}^{\ell }y_{\ell })$ are indeed convex and satisfy (\ref{11}).

Next we evaluate the blocks of the profile $U^{cut}$ in the same fashion.
Agents in $N_{k}$ load exclusively $A_{k}$ therefore if $z$ implement $%
U^{cut}$ we have $z_{A_{k}}=\frac{n_{k}}{n}$ and $U_{NA_{k}}^{cut}=\frac{%
n_{k}}{n}p_{k}$. We can find as above convex weights $\theta _{\ell
}^{k},1\leq \ell \leq k,$ such that%
\[
U_{N_{\ell }A_{k}}^{cut}=\theta _{\ell }^{k}\frac{n_{k}}{n}p_{k}\text{ for
all }1\leq \ell \leq k\leq K 
\]%
and then as above we get%
\[
U_{N_{k}}^{cut}=\sum_{\ell =k}^{K}\theta _{k}^{\ell }\frac{n_{\ell }}{n}%
p_{\ell } . 
\]

Assume now the profile $U^{cut}$ is $\varepsilon $-efficient:\ $U^{cut}\leq
\varepsilon U$ for some feasible $U$. From (\ref{11}) we find convex weights 
$\pi $ such that $U_{N_{k}}^{cut}\leq \varepsilon \pi _{k}p_{k}$ for all $k$%
, which implies%
\[
\varepsilon \geq \sum_{k=1}^{K}\frac{1}{p_{k}}U_{N_{k}}^{cut}=\sum_{\ell
=1}^{K}\frac{n_{\ell }}{n}\sum_{k=1}^{\ell }\theta _{k}^{\ell }\frac{p_{\ell
}}{p_{k}}. 
\]

The key inequality is $U_{N_{k}A_{k}}^{cut}\geq \frac{n_{k}}{n}$ because
agent $i\in N_{k}$ loads only $A_{k}$ containing his like-set: this implies $%
\theta _{k}^{k}\geq \frac{1}{p_{k}}$. Moreover in the sum $\sum_{k=1}^{\ell
}\theta _{k}^{\ell }\frac{p_{\ell }}{p_{k}}$ the terms $\frac{p_{\ell }}{%
p_{k}}$ increase in $k$. Combining these two observations we have for any $%
\ell \geq 2$:%
\[
\sum_{k=1}^{\ell }\theta _{k}^{\ell }\frac{p_{\ell }}{p_{k}}\geq {\Large (}%
\sum_{k=1}^{\ell -1}\theta _{k}^{\ell }{\Large )}\frac{p_{\ell }}{p_{1}}%
+\theta _{\ell }^{\ell }\geq (1-\frac{1}{p_{\ell }})\frac{p_{\ell }}{p_{1}}+%
\frac{1}{p_{\ell }}=\frac{p_{\ell }-1}{p_{1}}+\frac{1}{p_{\ell }}. 
\]%
We invoke now the inequality $\frac{\alpha -1}{p_{1}}+\frac{1}{\alpha }\geq 
\frac{2}{\sqrt{p_{1}}}-\frac{1}{p_{1}}$, for any $\alpha >0$, that we apply
to each $\alpha =p_{\ell },\ell \geq 2$, and combine with the two
inequalities above as well as $\theta _{1}^{1}=1$:%
\[
\varepsilon \geq \frac{n_{1}}{n}+(1-\frac{n_{1}}{n})(\frac{2}{\sqrt{p_{1}}}-%
\frac{1}{p_{1}}). 
\]%
Finally the term $\frac{2}{\sqrt{p_{1}}}-\frac{1}{p_{1}}$ decreases in $%
p_{1} $ and we know $p_{1}\leq n_{1}$, so we get%
\[
\varepsilon \geq \frac{1}{n}(n_{1}+(n-n_{1})(\frac{2}{\sqrt{n_{1}}}-\frac{1}{%
n_{1}})). 
\]%
It remains to compute the minimum of the above expression for fixed $n$ and
variable $n_{1}\in \lbrack 1,n]$. With the real variable $x$ instead of $%
n_{1}$ the right hand term and its derivative are%
\[
\varphi (x)=\frac{1}{n}(1+x-2\sqrt{x})+(\frac{2}{\sqrt{x}}-\frac{1}{x}%
)\Longrightarrow \varphi ^{\prime }(x)=(1-\frac{1}{\sqrt{x}})(\frac{1}{n}-%
\frac{1}{x^{\frac{3}{2}}}). 
\]%
therefore $x=n^{\frac{2}{3}}$ achieves the minimum and we compute%
\[
\varepsilon \geq \varphi (n^{\frac{2}{3}})=\frac{1}{n}+(1-\frac{1}{n^{\frac{1%
}{3}}})\frac{3}{n^{\frac{1}{3}}}. 
\]%
which is inequality (\ref{13}).\smallskip

\textit{Step 2: Lower bounding the worst case inefficiency of RP}

Fix $N$ and integers $k,d,\ell $ such that $n=kd$ and $2\leq \ell <k$. Fix a
partition $N^{1}\cup \cdots \cup N^{d}$ of $N$ where each subset contains $k$
agents. This construction requires $n\geq 6$ and is not feasible for all $n$.

We consider the problem with $A=D\cup C$ where $D=\{1,\cdots ,d\}$ and each $%
\delta \in D$ is liked exactly by the $k$ agents in $N^{\delta }$; also $|C|=%
\binom{n}{\ell }$ and each outcome in $C$ is liked exactly by a different
subset of $\ell $ agents.

The symmetric (egalitarian) and efficient outcome is the uniform
distribution in $D$ and yields the utility profile $U_{i}^{\ast }=\frac{1}{d}
$ for all $i$. We compute now the symmetric profile $U$ implemented by RP.

Fix an ordering $\sigma \in \Theta (N)$ and let $L$ be the set of its $\ell $
highest priority agents. In the resulting profile $U^{\sigma }$, the first $%
\ell $ agents have full utility (because there is $a\in C$ where they all
do). Two cases arise. In the favourable case $L$ is contained in some set $%
N^{\delta }$: then $\delta $ is the only efficient pure outcome liked by all
agents in $L$, thus it must be chosen by the $\sigma $-priority rule and $%
U_{N}^{\sigma }=k$. In the unfavourable case $L$ straddles two or more sets $%
N^{\delta }$ and there is only one outcome (in $C$) that everyone in $L$
like, so that $U_{N}^{\sigma }=\ell $. Therefore%
\[
U_{N}=\frac{d\binom{k}{\ell }}{\binom{n}{\ell }}\cdot k+{\Large (}1-\frac{d%
\binom{k}{\ell }}{\binom{n}{\ell }}k{\Large )}\cdot \ell =(k-\ell )\frac{n}{k%
}\frac{\binom{k}{\ell }}{\binom{n}{\ell }}+\ell .
\]%
\begin{equation}
\Longrightarrow \varepsilon (n)\leq \frac{U_{N}}{U_{N}^{\ast }}=(1-\frac{%
\ell }{k})\frac{(k-1)\cdots (k-\ell +1)}{(n-1)\cdots (n-\ell +1)}+\frac{\ell 
}{k}  \label{14}
\end{equation}%
For the asymptotic statement we use the inequality $\frac{\binom{k}{\ell }}{%
\binom{n}{\ell }}\leq (\frac{k}{n})^{\ell }$ and compute%
\[
\Rightarrow \frac{U_{i}}{U_{i}^{\prime }}=\frac{U_{N}}{U_{N}^{\prime }}\leq (%
\frac{k}{n})^{\ell -1}+\frac{\ell }{k}.
\]%
Then we choose $k\simeq \frac{n}{e}$ and $\ell \simeq \ln (n)$ so that $(%
\frac{k}{n})^{\ell -1}+\frac{\ell }{k}\simeq e\frac{\ln (n)}{n}$. The
systematic inequality $\varepsilon ^{rp}(n)\leq 6\frac{\ln (n)}{n}$ is
obtained by numerical estimations of (\ref{14}), omitted for brevity.$%
\blacksquare \smallskip $

\textit{Remark 2} \textit{The proof of Step 2 improves upon, with a similar
proof technique, Example 1 in \citep{BMS02a} establishing that RP is }$\frac{%
2}{\sqrt{n}}$\textit{inefficient.}

\section{Efficiency and Fairness; the Nash Max Product rule}

Our last rule of interest is a familiar compromise between the Utilitarian
and Egalitarian rules:

\[
\text{\textbf{Nash Max Product rule (NMP): }}F^{nsh}(M)=\arg \max_{U\in \Phi
(M)}\sum_{i\in N}\ln U_{i}. 
\]%
This rule is well defined because it solves a strictly convex program, and
obviously efficient.

Recall that Unanimity Fair Share offers welfare guarantees only to
coalitions of agents with identical preferences (clones). The first of our
two new \textquotedblleft Fair Share\textquotedblright\ axioms applies, much
more generally, to any group who can find at least one outcome that everyone
likes:

\[
\text{\textbf{Average Fair Share }(AFS)} 
\]%
\[
\text{for all }S\subseteq N\text{: }\{\exists a\in A:u_{ia}=1\text{ for all }%
i\in S\}\Longrightarrow \frac{1}{|S|}U_{S}\geq \frac{|S|}{n} . 
\]%
The next property conveys the idea that, as each agent is endowed with $1/n$%
-th of total decision power, any coalition of size $s$ can cumulate these
shares and impose that a mixture of their choice be chosen with probability
at least $\frac{s}{n}$:%
\[
\text{\textbf{Core Fair Share }(CFS)} 
\]%
\[
\text{for all }S\subseteq N:\nexists z\in \Delta (A)\text{ s. t. }\forall
i\in S,\text{ }U_{i}\leq \frac{|S|}{n}(u_{i}\cdot z)\text{ and }\exists
i,U_{i}<\frac{|S|}{n}(u_{i}\cdot z). . 
\]%
This is a familiar core stability property.

That UFS follows from either AFS or CFS is clear because we only consider
anonymous rules. Applying CFS to $S=N$ implies that the rule is efficient,
therefore neither the CUT or the RP rule meets CFS. In the example (\ref{3})
it happens that the AFS property selects uniquely the Nash mixture,\footnote{%
We leave the proof to the reader and give a similar example in the next
paragraph.} therefore CUT and RP fail AFS as well.

We illustrate the bite of AFS in the following example%
\[
\begin{array}{c|cccc}
& a & b & c & d \\ 
1 & 1 & 0 & 0 & 0 \\ 
2 & 1 & 1 & 1 & 0 \\ 
3 & 0 & 0 & 1 & 1 \\ 
4 & 0 & 1 & 0 & 1 \\ 
5 & 0 & 0 & 0 & 1%
\end{array}%
\]%
For a mixture $z=(\alpha ,\beta ,\gamma ,\delta )$ we apply AFS to $%
S=\{1,2\} $ and $T=\{3,4,5\}$%
\[
2\alpha +\beta +\gamma \geq \frac{4}{5}\text{ ; }\beta +\gamma +3\delta \geq 
\frac{9}{5}. 
\]%
Adding both inequalities gives $\delta \geq \frac{3}{5}$, then the first one
implies $\alpha \geq \frac{2}{5}$ so that $z=(\frac{2}{5},0,0,\frac{3}{5})$:
this is precisely the mixture selected by the NMP rule. Clearly $z$ meets
CFS as well.

By contrast Core Fair Share selects many more outcomes than the Nash
mixture, for instance $z^{\prime }=(\frac{1}{5},\frac{3}{20},\frac{3}{20},%
\frac{1}{2})$. So in this example AFS is more demanding than CFS, but in
general the two axioms are not logically related. For instance in the problem%
\[
\begin{array}{c|ccc}
1 & 1 & 0 & 0 \\ 
2 & 1 & 1 & 0 \\ 
3 & 0 & 1 & 0 \\ 
4 & 0 & 0 & 1%
\end{array}%
\]%
the mixture $z=(\frac{7}{20},\frac{7}{20},\frac{3}{10})$ gives the profile $%
U=(\frac{7}{20},\frac{7}{10},\frac{7}{20},\frac{3}{10})$. It passes the AFS
test but fails CFS because coalition $\{1,2,3\}$ achieves utilities $(\frac{3%
}{8},\frac{3}{4},\frac{3}{8})$ by implementing $\frac{3}{4}$ of the mixture $%
z^{\prime }=(\frac{1}{2},\frac{1}{2},0)$, and they all improve
strictly.\smallskip

\textbf{Theorem 3}

\noindent $i)$ \textit{The NMP rule is efficient and meets Strict
Participation, Average Fair Share, and Core Fair Share.}

\noindent $ii)$ \textit{The NMP rule is not Excludable
Strategyproof.\smallskip }

\textbf{Proof}

\noindent \textit{Step 1: We prove AFS and CFS}.

The separation inequality capturing the optimality of the Nash utility
profile $U^{\ast }=F^{nsh}(M)$ at problem $M$ writes as follows:%
\begin{equation}
\sum_{i\in N}\frac{U_{i}}{U_{i}^{\ast }}\leq \sum_{i\in N}\frac{U_{i}^{\ast }%
}{U_{i}^{\ast }}=n\text{ \ for all }U\in \Phi (M)  \label{6}
\end{equation}%
Fix $S\subseteq N$ and combine (\ref{6}) with Cauchy's inequality as follows%
\[
nU_{S}^{\ast }\geq {\large (}\sum_{i\in S}\frac{U_{i}}{U_{i}^{\ast }}{\large %
){\Large .}(}\sum_{i\in S}U_{i}^{\ast }{\large )}{\Large \geq }{\large (}%
\sum_{i\in S}\sqrt{U_{i}}{\large )}^{2}\Longrightarrow 
\]%
\begin{equation}
U_{S}^{\ast }\geq \frac{1}{n}\max_{U\in \Phi (M)}{\large (}\sum_{i\in S}%
\sqrt{U_{i}}{\large )}^{2}  \label{7}
\end{equation}%
The AFS property follows, because if there is some $a\in A$ such that $%
u_{ia}=1$ for all $i\in S$, the maximum on the right hand side is $|S|^{2}$.
To check CFS we assume there is a mixture $z$ such that $U_{i}^{\ast }\leq 
\frac{|S|}{n}(u_{i}\cdot z)$ for all $i\in S$ and use again (\ref{6}) to
compute:%
\[
n\geq \sum_{i\in S}\frac{u_{i}\cdot z}{U_{i}^{\ast }}\geq \frac{n}{|S|}%
\sum_{i\in S}\frac{U_{i}^{\ast }}{U_{i}^{\ast }}=n 
\]%
therefore none of the inequalities $U_{i}^{\ast }\leq \frac{|S|}{n}%
(u_{i}\cdot z)$ can be strict.\smallskip

\noindent \textit{Step 2}: \textit{We check PART}$^{\ast }$.

In a preliminary result we fix $S\subset 
\mathbb{R}
_{+}^{N}$ convex and compact, and write $S(-1)$ for its projection on $%
\mathbb{R}
_{+}^{N\diagdown 1}$. Define%
\[
U^{\ast }=\arg \max_{U\in S}\sum_{i\in N}\ln (U_{i}) 
\]%
\[
\overline{U}_{-1}=\arg \max_{U_{-1}\in S(-1)}\sum_{i\in N\diagdown 1}\ln
(U_{i})\text{ and }\overline{U}_{1}=\max_{(U_{1},\overline{U}_{-1})\in
S}U_{1} . 
\]%
Inequality $U_{1}^{\ast }<\overline{U}_{1}$ brings a contradiction as follows%
\[
\sum_{i\in N}\ln (\overline{U}_{i})\geq \ln (\overline{U}_{1})+\sum_{i\in
N\diagdown \{1\}}\ln (U_{i}^{\ast })>\sum_{i\in N}\ln (U_{i}^{\ast }) . 
\]%
Assume next $U_{1}^{\ast }=\overline{U}_{1}$. The right hand inequality
above becomes an equality, so we get $\sum_{i\in N}\ln (\overline{U}%
_{i})=\sum_{i\in N}\ln (U_{i}^{\ast })$ and finally $\overline{U}=U^{\ast }$%
. Summing up, we have just proven:%
\begin{equation}
U_{1}^{\ast }\geq \overline{U}_{1}\text{\textit{; and if }}U_{1}^{\ast }=%
\overline{U}_{1}\text{\textit{\ then }}U_{-1}^{\ast }=\overline{U}_{-1}
\label{2}
\end{equation}

Applying (\ref{2}) to $S=\Phi (M)$, $U^{\ast }=F^{nsh}(M)$, $\overline{U}%
_{-1}=F^{nsh}(M(-1))$ gives $\overline{U}_{1}=U_{1}(-1)$ and $U_{1}^{\ast
}\geq \overline{U}_{1}$, the first inequality in PART$^{\ast }$ (i.e.,
PART). To check the second we can assume that any two columns of $u$ are
different, for if two columns are identical one of them can be eliminated as
a redundant outcome. Also recall that no row of $u$ is null.

Because $U_{i}^{\ast }>0$ for all $i$, the statement is true if $\overline{U}%
_{1}=0$. We assume now $0<U_{1}^{\ast }=\overline{U}_{1}<1$ and derive a
contradiction. Property (\ref{2}) implies $U^{\ast }=\overline{U}$,
therefore there is some $z\in \Delta (A)$ solving both problems: $z\in
f^{nsh}(M)\cap f^{nsh}(M\mathcal{(}-1))$.

As $0<U_{1}^{\ast }<1$ the mixture $z$ cannot be deterministic, moreover
there exists two outcomes $a,b$ in the support $[z]$ of $z$ such that $%
u_{1a}=1,u_{1b}=0$. Writing $N(x;y)$ for the set of agents in $N$ who like $%
x $ and dislike $y$, this means $1\in N(a;b)$.

Note that $N(b;a)$ must contain at least one $i\in N\diagdown 1$: otherwise
the column $U_{a}$ dominates column $U_{b}$ (outcome $b$ is Pareto inferior
to $a$) which contradicts the efficiency of $z$ in $M$. We claim that $%
N(a;b) $ as well contains some $j\in N\diagdown 1$: suppose not, then the
restriction of column $U_{b}$ to $N\diagdown 1$ either dominates the
corresponding restriction of $U_{a}$, or these two restricted columns are
equal; the former case contradicts efficiency of $z$ in $M(-1)$, the latter
contradicts its efficiency in $M$.

We have shown that $N(a;b)$ and $N(b;a)$ both contains at least one outcome
in $N\diagdown 1$. Recalling that $z_{a},z_{b}$ are both positive, we define 
$z(\varepsilon )$ by shifting the weight $\varepsilon $ from $a$ to $b$:
this outcome is well defined for $\varepsilon $ small enough and of
arbitrary sign; such a shift does not affect agents outside $N(a;b)\cup
N(b;a)$. From $z\in f^{nsh}(M\mathcal{(}-1))$ we see that the strictly
concave function%
\[
\varphi (\varepsilon )=\sum_{i\in (N(a;b)\cup N(b;a))\diagdown 1}\ln
(u_{i}\cdot z(\varepsilon )) . 
\]%
reaches its maximum at $\varepsilon =0$. And $z\in f^{nsh}(M)$ implies that
the function $\varphi (\varepsilon )+\ln (u_{1}\cdot z(\varepsilon ))$ is
also maximal at $\varepsilon =0$: this is a contradiction because $\ln
(u_{1}\cdot z(\varepsilon ))$ decreases strictly in $\varepsilon $.$%
\blacksquare \smallskip $

The proof that the NMP rule fails EXSP is more involved and relegated to the
Appendix. There we construct an example with $|A|=4$ and $n=860$ where it
violates the SP$^{+}$ property. We also report a computer generated example
with 36 agents proving the same point. This prompts the following open
question: \textit{what are the smallest sizes of }$|N|,|A|$\textit{\
ensuring that NMP violates EXSP ?}\smallskip

\textit{Remark 3.} Another version of the group fair share requirement is
proposed by \citet{BMS02a}. {The same concept was independently proposed by %
\citet{Dudd15a} who referred to simply as \emph{proportional share}~%
\citep{Dudd15a}. For the sake of consistency with our other notions, we will
refer to it as \emph{Group Fair Share}(GFS).}

Writing $u^{\ast S}$ for the maximum of all utility functions in $S$ ($%
u_{a}^{\ast S}=\max_{i\in S}u_{ia}$), this condition is%
\[
U^{\ast S}\geq \frac{|S|}{n}\text{ for all }S . 
\]%
It is clearly stronger than UFS, but strictly weaker than CFS. {Both CUT and
RP satisfy GFS.\smallskip }

\textit{Remark 4. }{It has been mentioned as an open problem, in the more
general voting model with vNM-preferences, whether there exists some rule
that satisfies Very Strong Stochastic Dominance Participation and Stochastic
Dominance Efficiency for weak orders~\citep{BBH15b,Bran17a}. Because NMP
satisfies both Strict Participation and Efficiency, we see that this
question is resolved at least for the case of dichotomous preferences.}

\section{Decentralization}

We introduce a \textit{Decentralization }(DEC) property for polarized
societies\textit{. }Say the agents and the deterministic outcomes are
color-coded with the same set of colors: we call a profile of preferences 
\textit{polarized }if each agent only likes outcomes of his own color. The
requirement is that if I am red, the number of green agents will matter to
me but not their preferences\textit{\ }inside green outcomes. This natural
independence property adds to the appeal of the NMP rule, but also of the
CUT and RP rules.

Consider a problem $M=(N,A,u)$ and two partitions $\Gamma =(N^{k})_{k=1}^{K}$
and $\Lambda =(A^{k})_{k=1}^{K}$ of $N$ and $A$ respectively. We call this
problem \textit{polarized along the partitions} $\Gamma ,\Lambda $ if $%
u_{ia}=0$ whenever $i\in N^{k},a\in A^{k}$, and $k\neq k^{\prime }$. Then if 
$u^{k}$ is the restriction of $u$ to $N^{k}\times A^{k}$, problem $M$ is
captured by its $K$ \textit{subproblems} $M^{k}=(N^{k},A^{k},u^{k})$. We
write $\Pi (\Gamma ,\Lambda )$ the set of polarized problems.%
\[
\text{\textbf{Decentralization }(DEC): for any }\Gamma ,\Lambda \text{ and }%
k 
\]%
\[
\{M,M^{\prime }\in \Pi (\Gamma ,\Lambda )\text{ and }u_{ia}=u_{ia}^{\prime }%
\text{ if }i\in N^{k},a\in A^{k}\}\Longrightarrow F_{i}(M)=F_{i}(M^{\prime })%
\text{ for }i\in N^{k} . 
\]%
Combined with the UFS property, this implies that in a polarized problem,
each colored subset $N^{k}$ chooses the distribution in $\Delta (A^{k})$ as
if other colors were not present, then the selected outcome in $f(M^{k})$ is
weighted down in proportion of the size of $N^{k}$.

\textbf{Proposition\label{decentralize1} }\textit{The Nash, Conditional
Utilitarian, and Random Priority rules meet Decentralization. Moreover for
any polarized problem }$M\in \Pi (\Gamma ,\Lambda )$ \textit{they satisfy}%
\begin{equation}
F(M)=\sum_{k=1}^{K}\frac{|N^{k}|}{n}F(M^{k})  \label{12}
\end{equation}%
\textit{where the profile }$F(M^{k})$\textit{\ is filled with zeros outside }%
$M^{k}$.\textit{\smallskip }

\noindent Check that the Utilitarian and Egalitarian rules violate DEC.
Consider the two polarized problems along the partition $\{1\}\cup (2,3\}$:%
\[
M:%
\begin{array}{cccc}
1 & 1 & 0 & 0 \\ 
2 & 0 & 1 & 0 \\ 
3 & 0 & 0 & 1%
\end{array}%
\text{ \ \ }M^{\prime }:%
\begin{array}{cccc}
1 & 1 & 0 & 0 \\ 
2 & 0 & 1 & 0 \\ 
3 & 0 & 1 & 1%
\end{array}%
. 
\]%
Both UTIL and EGAL choose $z=(\frac{1}{3},\frac{1}{3},\frac{1}{3})$ at $M$,
but at $M^{\prime }$ they pick respectively $z^{\prime }=(0,1,0)$ and $%
z^{\prime \prime }=(\frac{1}{2},\frac{1}{2},0)$, in contradiction of DEC.

\textbf{Proof of Proposition}%
\index{decentralize1}

\noindent \textit{Step 1.} \textit{NMP meets DEC}

We prove it for two partitions $\Gamma =(N_{1},N_{2})$ and $\Lambda
=(A_{1},A_{2})$ as the general case with arbitrary $K$ is just as easy. We
fix a problem $M\in \Pi (\Gamma ,\Lambda )$ with the two subproblems $%
M^{k}=(N^{k},A^{k},u^{k})$, $k=1,2$: there is no null row in $u^{k}$ because
there is none in the grand matrix $u$. Let $z^{\ast }\in f^{N}(M)$ be (one
of) NMP's choice in $M$. By IFS both weights $z_{A_{1}}^{\ast }=\lambda _{1}$
and $z_{A_{2}}^{\ast }=\lambda _{2}$ are strictly positive (and sum to $1$).
By definition (\ref{4}) the restriction $z^{\ast }(A^{k})$ of $z^{\ast }$ to 
$A^{k}$ solves%
\[
\max_{%
\widetilde{z}\geq 0,\widetilde{z}_{A^{k}}=\lambda _{k}}\sum_{i\in N^{k}}\ln
(u_{i}\cdot \widetilde{z}) . 
\]%
Changing the variables $\widetilde{z}$ to $z=\frac{1}{\lambda _{k}}%
\widetilde{z}$, we see that $\frac{1}{\lambda _{k}}z^{\ast }(A^{k})$ solves%
\[
\max_{z\in \Delta (A^{k})}\sum_{i\in N^{k}}\ln (u_{i}\cdot \lambda
z)=n_{k}\ln \lambda _{k}+\max_{z\in \Delta (A^{k})}\sum_{i\in N^{k}}\ln
(u_{i}\cdot z) 
\]%
therefore $z^{\ast }(A^{k})=\lambda _{k}z^{k}$, where $z^{k}\in f^{N}(M^{k})$%
, and%
\[
\max_{\widetilde{z}\geq 0,\widetilde{z}_{A^{k}}=\lambda _{k}}\sum_{i\in
N^{k}}\ln (u_{i}\cdot \widetilde{z})=n_{k}\ln \lambda _{k}+\sum_{i\in
N^{k}}\ln F_{i}^{N}(M^{k}) . 
\]%
It is now clear that the optimal choice of $\lambda _{1},\lambda _{2}$ is%
\[
(\lambda _{1},\lambda _{2})=\arg \max_{\lambda _{1}+\lambda
_{2}=1}\{n_{1}\ln \lambda _{1}+n_{2}\ln \lambda _{2}\}=(\frac{n_{1}}{n},%
\frac{n_{2}}{n}) . 
\]%
In particular $\lambda $ depends only on the sizes of the partition sets,
and $U_{i}^{\ast }=\frac{n_{k}}{n}F_{i}^{N}(M^{k})$ for each $i\in N^{k}$,
the desired property (\ref{12}).\smallskip

\noindent \textit{Step 2 CUT and RP meet DEC}

Fix $M\in \Pi (\Gamma ,\Lambda )$ some $k,1\leq k\leq K$ and some $i\in
N^{k} $. Under the CUT rule only the agents in $N^{k}$ will load the
outcomes that $i$ likes, and they will do it exactly as in the problem $%
M^{k} $, except that each $j\in N^{k}$ will spread a total weight of $\frac{1%
}{n}$ instead of $\frac{1}{n_{k}}$. This implies (\ref{12}). The proof for
RP is just as easy. $\blacksquare $

\section{Computation}

In this section we first discuss the computational aspects of the rules we
have considered in the paper. We then report on some experiments where we
examined the utilitarian performance of the four rules, which in turn gives
a lower bound on their efficiency.\smallskip

\textit{1. Computational complexity}

The CUT rule is the easiest to compute of the four. In the like-set of each
agent we simply need to identify those liked by the largest number of other
agents.

For EGAL, the outcome can be computed in polynomial-time by solving at most $%
n+1$ linear programs each with $|A|$ variables. The algorithm was presented
by \citet{AzSt14a}.

The RP outcome is \#P-complete to compute even under dichotomous preferences~%
\citep{ABB13b}. Therefore unless P=NP, it is unlikely that there exists an
efficient algorithm for computing the RP outcome. For RP, it is even open
whether there exists an FPRAS (Fully Polynomial-time Approximation Scheme)
for computing the outcome shares/probabilities.

As for NMP, in contrast to RP, an approximate solution can be computed
relatively fast by using general optimisation packages and solvers. The
problem is to maximize a convex objective $\sum_{i\in N}\log (u_{i}\cdot z)$
where $z$ is a feasible mixture. \citet{BMS05a} discussed some standard
approaches to approximate the solution. 
\smallskip

\textit{2. Experiments}

We ran some experiments for small numbers of agents and outcomes. It is
difficult to evaluate in a given problem the degree of inefficiency of a
given mixture\ $z$ as in Definition 1, $iii)$. However the ratio of
utilitarian welfare at $z$ to the maximum utilitarian welfare gives a lower
bound on $\varepsilon $, and it is much easier to compute.

For each combination of $n$ and $|A|$ in $\{3,5,7,10,15,20\}$ and for each
rule, we examined under the impartial culture (1) the minimum of this ratio
and (2) its average. The results are listed in Tables 1--8. For RP, we did
not run the experiments for $n=15$ and $20$ because the computation becomes
very slow. This illustrates the computational infeasibility of RP when we
want the exact mixture, even for a relatively modest number of agents.

As the number of agents increase, the ratios start to get worse. But for a
fixed number of agents, the ratios do not necessarily get worse as we
increase the number of alternatives. We note that CUT seems to fare
marginally but consistently better than NMP, RP, and EGAL in the utilitarian
metric. This is especially so when we consider the average rather than the
worst ratios.

We note that NMP rule's fairness constraints also lead to loss of
utilitarian welfare. \citet{FGM16a} show that on certain real-world
participatory budgeting datasets, core fair outcomes often coincide with
welfare maximizing ones. Since the objective of EGAL is diametrically
opposed to utilitarian objectives, it is not surprising that EGAL fares the
worst in the utilitarian metric among the rules we consider. In particular
its worst case ratios drop rapidly as we increase the number of agents and
outcomes.

\begin{table}[t!]
\label{table:rp} 
\begin{tabular}[h]{ccccccccccccc}
\hline
$|N|\downarrow |A|\rightarrow$ & 3 & 5 & 7 & 10 & 15 & 20\\
3 & 0.8314 & 0.8155 & 0.8069 & 0.8005 & 0.781 & 0.7149 \\ 
5 & 0.7777 & 0.7778 & 0.7322 & 0.7531 & 0.7072 & 0.7172  \\ 
7 & 0.7678 & 0.80790 & 0.7373 & 0.695 & 0.7581 & 0.7109  \\ 
10 & 0.7524 & 0.7334 & 0.808 & 0.7843 & 0.7857 & 0.7204 \\ 
15 & 0.7862 & 0.8029 & 0.7561 & 0.7801 & 0.7747 & 0.7737
\\ 
20 & 0.792 & 0.8234 & 0.7764 & 0.8155 & 0.7505 & 0.7896  \\ 
\hline
\end{tabular}%
\caption{Minimum ratio of utilitarian welfare under the NMP rule to maximum
utilitarian welfare for 100 profiles draws under impartial culture
assumption for each combination of \# agents and \# outcomes. }
\label{table:summary:random}
\end{table}

\begin{table}[t!]
\label{table:rp} 
\begin{tabular}[h]{ccccccccccccc}
\hline
$|N|\downarrow |A|\rightarrow$ & 3 & 5 & 7 & 10 & 15 & 20\\
3 & 0.9451 & 0.9652 & 0.9722 & 0.9678 & 0.9759 & 0.9634 \\
5 & 0.9171 & 0.9309 & 0.9421 & 0.9377 & 0.9335 & 0.9004 \\
7 & 0.8926 & 0.9324 & 0.9171 & 0.9277 & 0.9121 & 0.8856 \\
10 & 0.8921 & 0.9014 & 0.91 & 0.9094 & 0.9056 & 0.8873 \\
15 & 0.893 & 0.9013 & 0.8911 & 0.9049 & 0.8984 & 0.8774 \\
20 & 0.8948 & 0.9001 & 0.8909 & 0.9047 & 0.9049 & 0.8941
\\ \hline
\end{tabular}%
\caption{Average ratio of utilitarian welfare the NMP rule to maximum
utilitarian welfare for 100 profiles draws under impartial culture
assumption for each combination of \# agents and \# outcomes. }
\label{table:summary:random}
\end{table}

\begin{table}[t!]
\label{table:rp} 
\begin{tabular}[h]{ccccccccccccc}
\hline
$|N|\downarrow |A|\rightarrow$ & 3 & 5 & 7 & 10 & 15 & 20 \\
3 & 0.75 & 0.6397 & 0.5333 & 0.4815 & 0.4333 & 0.3743 \\
5 & 0.625 & 0.3919 & 0.4244 & 0.4592 & 0.4956 & 0.403  \\
7 & 0.5833 & 0.492 & 0.3632 & 0.5102 & 0.5599 & 0.5799 \\
10 & 0.5834 & 0.375 & 0.4952 & 0.4253 & 0.5689 & 0.5696  \\
15 & 0.5129 & 0.5525 & 0.57 & 0.4361 & 0.5198 & 0.5817 \\
20 & 0.6001 & 0.625 & 0.5927 & 0.5525 & 0.6425 & 0.5656  \\
\hline
\end{tabular}%
\caption{Minimum ratio of utilitarian welfare under EGAL to maximum
utilitarian welfare for 100 profiles draws under impartial culture
assumption for each combination of \# agents and \# outcomes. }
\label{table:summary:random}
\end{table}

\begin{table}[t!]
\label{table:rp} 
\begin{tabular}[h]{ccccccccccccc}
\hline
$|N|\downarrow |A|\rightarrow$ & 3 & 5 & 7 & 10 & 15 & 20 \\
3 & 0.9325 & 0.9256 & 0.8838 & 0.8075 & 0.844 & 0.8408 \\
5 & 0.8482 & 0.8484 & 0.781 & 0.8019 & 0.82 & 0.8175 \\
7 & 0.8221 & 0.8131 & 0.7817 & 0.7978 & 0.7992 & 0.8118 \\
10 & 0.8176 & 0.8049 & 0.7902 & 0.7639 & 0.8152 & 0.7803
\\ 
15 & 0.8267 & 0.807 & 0.7805 & 0.7476 & 0.8259 & 0.8009 \\
20 & 0.8414 & 0.8278 & 0.8121 & 0.7748 & 0.8265 & 0.8084 
\\ \hline
\end{tabular}%
\caption{Average ratio of utilitarian welfare under EGAL to maximum
utilitarian welfare for 100 profiles draws under impartial culture
assumption for each combination of \# agents and \# outcomes. }
\label{table:summary:random}
\end{table}

\begin{table}[t!]
\label{table:rp} 
\begin{tabular}[h]{ccccccccccccc}
\hline
$|N|\downarrow |A|\rightarrow$ & 3 & 5 & 7 & 10 & 15 & 20 \\
3 & 0.8333 & 0.8333 & 0.8333 & 0.8333 & 0.8333 & 0.8333 \\
5 & 0.8 & 0.7333 & 0.8 & 0.8 & 0.8 & 0.8667 \\
7 & 0.75 & 0.7619 & 0.8571 & 0.8214 & 0.8857 & 0.8571 \\
10 & 0.8 & 0.8 & 0.8714 & 0.86 & 0.8667 & 0.8833 \\
15 & 0.8 & 0.8444 & 0.8583 & 0.8417 & 0.8741 & 0.8815 \\
20 & 0.8038 & 0.85 & 0.8773 & 0.9 & 0.8944 & 0.8727 \\
\hline
\end{tabular}%
\caption{Minimum ratio of utilitarian welfare under CUT to maximum
utilitarian welfare for 100 profiles draws under impartial culture
assumption for each combination of \# agents and \# outcomes. }
\label{table:summary:random}
\end{table}

\begin{table}[t!]
\label{table:rp} 
\begin{tabular}[h]{ccccccccccccc}
\hline
$|N|\downarrow |A|\rightarrow$ & 3 & 5 & 7 & 10 & 15 & 20 \\
3 & 0.9333 & 0.9717 & 0.9717 & 0.9867 & 0.9867 & 0.995 \\
5 & 0.9372 & 0.9452 & 0.959 & 0.9748 & 0.969 & 0.9757 \\
7 & 0.9139 & 0.9468 & 0.9549 & 0.9624 & 0.969 & 0.9778 \\
10 & 0.9194 & 0.9383 & 0.9502 & 0.9586 & 0.9576 & 0.965 \\
15 & 0.9263 & 0.9276 & 0.9483 & 0.9483 & 0.9567 & 0.9634
\\ 
20 & 0.9195 & 0.9332 & 0.9486 & 0.955 & 0.9588 & 0.9631 \\
\hline
\end{tabular}%
\caption{Average ratio of utilitarian welfare under CUT to maximum
utilitarian welfare for 100 profiles draws under impartial culture
assumption for each combination of \# agents and \# outcomes. }
\label{table:summary:random}
\end{table}

\begin{table}[t!]
\label{table:rp} 
\begin{tabular}[h]{ccccccccc}
\hline
$|N|\downarrow |A|\rightarrow$ & 3 & 5 & 7 & 10 & 15 & 20 \\ 
3 & 0.8333 & 0.8333 & 0.8333 & 0.8333 & 0.8333 & 0.8333 \\ 
5 & 0.7778 & 0.7 & 0.7778 & 0.7778 & 0.7 & 0.8 \\ 
7 & 0.7679 & 0.75 & 0.8036 & 0.75 & 0.7943 & 0.7778 \\ 
10 & 0.7778 & 0.7737 & 0.7596 & 0.8116 & 0.7684 & 0.8031 \\ \hline
\end{tabular}%
\caption{Minimum ratio of utilitarian welfare under RP to maximum
utilitarian welfare for 100 profiles draws under impartial culture
assumption for each combination of \# agents and \# outcomes. }
\label{table:summary:random}
\end{table}

\begin{table}[t!]
\label{table:rp} 
\begin{tabular}[h]{ccccccccccc}
\hline
$|N|\downarrow |A|\rightarrow$ & 3 & 5 & 7 & 10 & 15 & 20 \\ 
3 & 0.9483 & 0.9733 & 0.9883 & 0.99 & 0.9867 & 0.9933\\ 
5 & 0.8992 & 0.9302 & 0.9351 & 0.9471 & 0.9512 & 0.962 \\ 
7 & 0.8851 & 0.8952 & 0.9143 & 0.9182 & 0.929 & 0.9305\\ 
10 & 0.8839 & 0.89 & 0.8911 & 0.8969 & 0.9 & 0.8997  \\ \hline
\end{tabular}%
\caption{Average ratio of utilitarian welfare under RP to maximum
utilitarian welfare for 100 profiles draws under impartial culture
assumption for each combination of \# agents and \# outcomes. }
\label{table:summary:random}
\end{table}

%
%
%

\newpage

\newpage
\newpage
\newpage

\section{Conclusion and Open Questions}

1) {We compared the relative merits of some well-known rules (EGAL, RP, NMP)
and of an (essentially) new one (CUT), for the model of
probabilistic/fractional voting under dichotomous preferences. We did so by
taking a more nuanced and fine-grained approach to standard concepts such as
strategyproofness, participation incentives, and welfare guarantees, of
which we introduced new versions, both weaker and stronger than the existing
ones. Some of the results are summarised in Table~\ref{table:summary:dich}.}

\begin{table*}[h!]
\centering
\label{tab:compare} 
\begin{tabular}{lccccccc}
\hline
& RP & CUT & UTIL & EGAL & NMP \\ 
Properties &  &  &  &  &  \\ \hline
EFF (Efficiency) & -- & -- & + & + & + \\ \hline
EXSP = SP$^{-} \wedge \text{SP}^{+}$ (Excludable SP) & + & + & + & + & -- \\ 
SP = SP$^{\ast} \wedge \text{SP}^{+}$ (Strategyproofness) & + & + & + & -- & 
-- \\ \hline
IFS (Individual Fair Share) & + & + & -- & + & + \\ 
GFS (Group Fair Share) & + & + & -- & -- & + \\ 
AFS (Avg. Fair Share) & -- & -- & -- & -- & + \\ 
CFS (Core Fair Share) & -- & -- & -- & -- & + \\ \hline
PART (Participation) & + & + & + & + & {+} \\ 
PART$^{\ast}$ (Strict participation) & + & + & -- & -- & {+} \\ \hline
DEC (Decentralisation) & + & + & -- & -- & + \\ \hline
Known Polynomial-time Algorithm & -- & + & + & + & -- \\ \hline
\end{tabular}%
\caption{Properties satisfied by rules under dichotomous preferences.}
\label{table:summary:dich}
\end{table*}

{The two rules that are especially desirable in the instances where
protection of minorities and participation concerns matter most are CUT and
NMP. The \textit{Conditional Utilitarian} rule is strongly incentive
compatible, but in extreme cases it may be severely inefficient. The \textit{%
Nash Max Product} rule is efficient and gives much better guarantees to
groups of agents than CUT, but it fails even the weak form of
strategyproofness where outcomes are excludable.\smallskip\ }

2) Our results also identify two especially interesting open questions. We
know from \cite{BMS05a} that Efficiency, Individual Fair Share and
Strategyproofness are incompatible. If we are content to achieve only the
excludable version of Strategyproofness, this incompatibility disappears,
and the Egalitarian rule is an example. The unpalatable feature of this rule
is that it pays no attention to clones (subgroups of agents with identical
preferences) hence offers no protection to sizable minorities. But can a
rule combine Efficiency, Excludable Strategyproofness and Strict
Participation; or Efficiency, Excludable Strategyproofness and Unanimous
Fair Share? Such a rule would be a serious new contender in our fair mixing
model.\smallskip

3) \citet{BMS05a} defined, and \citet{BMS02a} studied, a family of welfarist
rules directly borrowed from classical social choice theory. Fix an
increasing, strictly concave, and continuous function $h$ on $[0.1]$. A rule
in the sense of Definition 1 is obtained by maximizing the sum of individual
utilities weighted by $h$:%
\begin{equation}
\mathbf{h}\text{\textbf{-rule: }}f(M)=\arg \max_{U\in \Phi (M)}\sum_{i\in
N}h(U_{i})  \label{4}
\end{equation}%
This maximization has a unique solution in $\Phi (M)$. The NMP rule is of
course a paramount example.

All $h$-rules are efficient, and by mimicking Step 2 in the proof of Theorem
3, we see that they satisfy PART$^{\ast }$ provided $h^{\prime }(0)=\infty $%
. They satisfy (resp. fail) IFS if $h$ is at least as concave as (resp. less
concave than) the $log$ function; but NMP is the only $h$-rule meeting UFS
(these two facts are already proven in \cite{BMS02a}). Finally all $h$-rules
fail EXSP and only NMP meets DEC. Thus they don't add much to our axiomatic
discussion.

However, once we observe that the EGAL and UTIL rules are the two end points
of the family of $h$-rules\footnote{%
When $h$ converges pointwise to a linear function, e.g. $h(x)=x^{q}$ with $%
q\uparrow 1$, the $h$-rule converges pointwise to UTIL; when $h$ becomes
infinitely concave, e.g. $h(x)=-x^{q}$ with $q\downarrow -\infty $, it
converges pointwise to EGAL.} the following intriguing facts emerges: most $%
h $-rules meet PART$^{\ast }$ but neither EGAL nor UTIL does; EGAL and UTIL
meet EXSP, but none of the $h$-rules does.

\section{Acknowledgements}

The authors thank Edward Lee for assistance with the code for some of the
rules. The comments of seminar participants at Seoul National University,,
Universit\'{e} Paris 1, and the Paris School of Economics are gratefully
acknowledged.

 \bibliographystyle{elsarticle-harv}

\begin{thebibliography}{40}
  \expandafter\ifx\csname natexlab\endcsname\relax\def\natexlab#1{#1}\fi
  \expandafter\ifx\csname url\endcsname\relax
    \def\url#1{\texttt{#1}}\fi
  \expandafter\ifx\csname urlprefix\endcsname\relax\def\urlprefix{URL }\fi

  \bibitem[{Abdulkadiro{\u{g}}lu and S{\"o}nmez(1998)}]{AbSo98a}
  Abdulkadiro{\u{g}}lu, A., S{\"o}nmez, T., 1998. Random serial dictatorship and
    the core from random endowments in house allocation problems. Econometrica
    66~(3), 689--701.

  \bibitem[{Aziz(2013)}]{Aziz13b}
  Aziz, H., 2013. {Maximal Recursive Rule: A New Social Decision Scheme}. In:
    Proceedings of the 23nd International Joint Conference on Artificial
    Intelligence (IJCAI). AAAI Press, pp. 34--40.

  \bibitem[{{A}ziz(2017)}]{Aziz17b}
  {A}ziz, H., 2017. A probabilistic approach to voting, allocation, matching, and
    coalition formation.

  \bibitem[{Aziz et~al.(2017{\natexlab{a}})Aziz, Brandl, Brandt, and
    Brill}]{ABBB15a}
  Aziz, H., Brandl, F., Brandt, F., Brill, M., 2017{\natexlab{a}}. On the
    tradeoff between efficiency and strategyproofnessWorking paper.

  \bibitem[{Aziz et~al.(2013)Aziz, Brandt, and Brill}]{ABB13b}
  Aziz, H., Brandt, F., Brill, M., 2013. The computational complexity of random
    serial dictatorship. Economics Letters 121~(3), 341--345.

  \bibitem[{Aziz et~al.(2017{\natexlab{b}})Aziz, Brill, Conitzer, Elkind,
    Freeman, and Walsh}]{ABC+16a}
  Aziz, H., Brill, M., Conitzer, V., Elkind, E., Freeman, R., Walsh, T.,
    2017{\natexlab{b}}. Justified representation in approval-based committee
    voting. Social Choice and Welfare 48~(2), 461--485.

  \bibitem[{Aziz and Stursberg(2014)}]{AzSt14a}
  Aziz, H., Stursberg, P., 2014. A generalization of probabilistic serial to
    randomized social choice. In: Proceedings of the 28th AAAI Conference on
    Artificial Intelligence (AAAI). AAAI Press, pp. 559--565.

  \bibitem[{Behrens(2017)}]{Behr17a}
  Behrens, J., 2017. The origins of liquid democracy. The Liquid Democracy
    Journal 5.

  \bibitem[{Benade et~al.(2017)Benade, Nath, Procaccia, and Shah}]{BNPS17a}
  Benade, G., Nath, S., Procaccia, A.~D., Shah, N., 2017. Preference elicitation
    for participatory budgeting. In: Proceedings of the 31st AAAI Conference on
    Artificial Intelligence (AAAI). AAAI Press, pp. 376--382.

  \bibitem[{Bogomolnaia and Moulin(2001)}]{BoMo01a}
  Bogomolnaia, A., Moulin, H., 2001. A new solution to the random assignment
    problem. Journal of Economic Theory 100~(2), 295--328.

  \bibitem[{Bogomolnaia and Moulin(2004)}]{BoMo04a}
  Bogomolnaia, A., Moulin, H., 2004. Random matching under dichotomous
    preferences. Econometrica 72~(1), 257--279.

  \bibitem[{Bogomolnaia et~al.(2017)Bogomolnaia, Moulin, Sandomirskyi, and
    Yanovskaya}]{BMSY17a}
  Bogomolnaia, A., Moulin, H., Sandomirskyi, F., Yanovskaya, E., 2017.
    Competitive division of a mixed manna. Econometrica.

  \bibitem[{Bogomolnaia et~al.(2002)Bogomolnaia, Moulin, and Stong}]{BMS02a}
  Bogomolnaia, A., Moulin, H., Stong, R., 2002. Collective choice under
    dichotomous preferences.

  \bibitem[{Bogomolnaia et~al.(2005)Bogomolnaia, Moulin, and Stong}]{BMS05a}
  Bogomolnaia, A., Moulin, H., Stong, R., 2005. Collective choice under
    dichotomous preferences. Journal of Economic Theory 122~(2), 165--184.

  \bibitem[{Brandl et~al.(2015)Brandl, Brandt, and Hofbauer}]{BBH15b}
  Brandl, F., Brandt, F., Hofbauer, J., 2015. Incentives for participation and
    abstention in probabilistic social choice. In: Proceedings of the 14th
    International Conference on Autonomous Agents and Multi-Agent Systems
    (AAMAS). IFAAMAS, pp. 1411--1419.

  \bibitem[{Brandl et~al.(2016)Brandl, Brandt, and Seedig}]{Bran13a}
  Brandl, F., Brandt, F., Seedig, H.~G., 2016. Consistent probabilistic social
    choice. Econometrica 84~(5), 1839--1880.

  \bibitem[{Brandt(2017)}]{Bran17a}
  Brandt, F., 2017. Rolling the dice: {R}ecent results in probabilistic social
    choice. In: Endriss, U. (Ed.), Trends in Computational Social Choice. AI
    Access, Ch.~1, pp. 3--26.

  \bibitem[{Brill(2017)}]{Bri17a}
  Brill, M., 2017. Interactive democracy: New challenges for social choice
    theory. In: Laslier, J., Moulin, H., Sanver, R., Zwicker, W. (Eds.), The
    Future of Economic Design. Springer.

  \bibitem[{Cabannes(2004)}]{Caba04a}
  Cabannes, Y., 2004. Participatory budgeting: a significant contribution to
    participatory democracy. Environment and Urbanization 16~(1), 27--46.

  \bibitem[{Caragiannis et~al.(2016)Caragiannis, Kurokawa, Moulin, Procaccia,
    Shah, and Wang}]{CKM+16a}
  Caragiannis, I., Kurokawa, D., Moulin, H., Procaccia, A.~D., Shah, N., Wang,
    J., 2016. {The Unreasonable Fairness of Maximum Nash Welfare}. In:
    Proceedings of the 17th ACM Conference on Economics and Computation (ACM-EC).
    pp. 305--322.

  \bibitem[{Duddy(2015)}]{Dudd15a}
  Duddy, C., 2015. Fair sharing under dichotomous preferences. Mathematical
    Social Sciences 73, 1--5.

  \bibitem[{Fain et~al.(2016)Fain, Goel, and Munagala}]{FGM16a}
  Fain, B., Goel, A., Munagala, K., 2016. The core of the participatory budgeting
    problem. In: Web and Internet Economics - 12th International Conference,
    {WINE} 2016, Montreal, Canada, December 11-14, 2016, Proceedings. pp.
    384--399.

  \bibitem[{Fishburn(1984)}]{Fish84a}
  Fishburn, P.~C., 1984. Probabilistic social choice based on simple voting
    comparisons. Review of Economic Studies 51~(4), 683--692.

  \bibitem[{Fishburn and Brams(1983)}]{BrFi83a}
  Fishburn, P.~C., Brams, S.~J., 1983. Paradoxes of preferential voting.
    Mathematics Magazine 56~(4), 207--214.

  \bibitem[{Gibbard(1977)}]{Gibb77a}
  Gibbard, A., 1977. Manipulation of schemes that mix voting with chance.
    Econometrica 45~(3), 665--681.

  \bibitem[{Gordon(1994)}]{Gord94a}
  Gordon, J., 1994. Institutions as relational investors: A new look at
    cumulative voting. Columbia Law Review 94~(1), 124--192.

  \bibitem[{Grandi(2017)}]{Gran17a}
  Grandi, U., 2017. Agent-mediated social choice. In: Laslier, J., Moulin, H.,
    Sanver, R., Zwicker, W. (Eds.), The Future of Economic Design. Springer.

  \bibitem[{Hughes and Sasse(2003)}]{HuSa03b}
  Hughes, J., Sasse, G., 2003. Monitoring the monitors : Eu enlargement
    conditionality and minority protection in the ceecs. Journal on ethnopolitics
    and minority issues in Europe 1, 35.

  \bibitem[{Laffond et~al.(1993)Laffond, Laslier, and {Le Breton}}]{LLL93b}
  Laffond, G., Laslier, J.-F., {Le Breton}, M., 1993. The bipartisan set of a
    tournament game. Games and Economic Behavior 5~(1), 182--201.

  \bibitem[{Moulin(1981)}]{Mou81a}
  Moulin, H., 1981. The proportional veto principle. Review of Economic Studies
    48~(3).

  \bibitem[{Moulin(1982)}]{Moul82a}
  Moulin, H., 1982. Voting with proportional veto power. Econometrica 50~(1),
    145--162.

  \bibitem[{Moulin(1988)}]{Moul88a}
  Moulin, H., 1988. Axioms of Cooperative Decision Making. Cambridge University
    Press.

  \bibitem[{Moulin(2003)}]{Moul03a}
  Moulin, H., 2003. Fair Division and Collective Welfare. The MIT Press.

  \bibitem[{Nash(1950)}]{Nash50b}
  Nash, J.~F., 1950. The bargaining problem. Econometrica 18~(2), 155--162.

  \bibitem[{Porta et~al.(2000)Porta, de~Silanes, Schleifer, and Vishny}]{PLSV00a}
  Porta, R.~L., de~Silanes, F.~L., Schleifer, A., Vishny, R., 2000. Investor
    protection and corporate governance. Journal of Financial Economics
    58~(1--2), 3--27.

  \bibitem[{Sawyer and MacRae(1962)}]{SaMa62a}
  Sawyer, J., MacRae, D., 1962. Game theory and cumulative voting in illinois:
    1902-1954. The American Political Science Review 56~(4), 936--946.

  \bibitem[{Steinhaus(1948)}]{Stei48a}
  Steinhaus, H., 1948. The problem of fair division. Econometrica 16, 101--104.

  \bibitem[{Thomson(2016)}]{Thom15a}
  Thomson, W., 2016. Introduction to the theory of fair allocation. In: Brandt,
    F., Conitzer, V., Endriss, U., Lang, J., Procaccia, A.~D. (Eds.), Handbook of
    Computational Social Choice. Cambridge University Press, Ch.~11.

  \bibitem[{Varian(1974)}]{Vari74a}
  Varian, H.~R., 1974. Equity, envy, and efficiency. Journal of Economic Theory
    9, 63--91.

  \bibitem[{Young(1950)}]{Youn50a}
  Young, G., 1950. The case for cumulative voting. Wisconsin Law Review 49--56.

  \end{thebibliography}

\section{Appendix}

\subsection{The NMP rule fails EXSP}

\subsubsection{A numerical example}

{Consider the following example with 36 agents and 4 outcomes. 
\[
\begin{array}{c|cccc}
& a & b & c & d \\ 
\text{No. of agents types} &  &  &  &  \\ \hline
4 & 1 & 0 & 0 & 0 \\ 
4 & 0 & 1 & 1 & 1 \\ 
1 & 0 & 0 & 1 & 0 \\ 
1 & 0 & 0 & 0 & 1 \\ 
2 & 1 & 1 & 1 & 0 \\ 
2 & 1 & 1 & 0 & 1 \\ 
7 & 1 & 0 & 1 & 0 \\ 
7 & 1 & 0 & 0 & 1 \\ 
4 & 0 & 1 & 1 & 0 \\ 
4 & 0 & 1 & 0 & 1 \\ 
&  &  &  & 
\end{array}%
\]%
The outcome of NMP is $(0.4163514575435199,$ $0.08787730532715962,$ $%
0.2479123840667547,$ $0.24785885306256383).$ If one agent of type one
additionally liked $b$, the profile is as follows. 
\[
\begin{array}{c|cccc}
& a & b & c & d \\ 
\text{No. of agents types} &  &  &  &  \\ \hline
3 & 1 & 0 & 0 & 0 \\ 
1 & 1 & 1 & 0 & 0 \\ 
4 & 0 & 1 & 1 & 1 \\ 
1 & 0 & 0 & 1 & 0 \\ 
1 & 0 & 0 & 0 & 1 \\ 
2 & 1 & 1 & 1 & 0 \\ 
2 & 1 & 1 & 0 & 1 \\ 
7 & 1 & 0 & 1 & 0 \\ 
7 & 1 & 0 & 0 & 1 \\ 
4 & 0 & 1 & 1 & 0 \\ 
4 & 0 & 1 & 0 & 1 \\ 
&  &  &  & 
\end{array}%
\]%
In this case, The outcome of NMP is $0.4179621510380684,$ $%
0.1389580435629242,$ $0.22150747720884034,$ $0.22157232819017458)$. Note
that the misreporting agent gets more utility (equivalently more probability
for outcome $a$ by additionally liking $b$).}

\subsubsection{A formal construction}

We fix $N$ and $A$ and describe a profile $u\in \Omega $ by the vector $%
(n_{S})_{S\in 2^{A}\diagdown \varnothing }$ of non negative integers, where $%
n_{S}$ is the number of agents $i$ with like-set $L_{i}=S$. Because $f^{nsh}$
is Anonymous, this is all we need to describe $f^{nsh}(u)$. If $z\in \Delta
(A)$ has full support ($z_{a}>0$ for all $a$), it is (uniquely) selected at $%
u$ if and only if the gradient of $z\rightarrow \varphi (z)=\sum_{N}\ln
(z_{L_{i}})$ is parallel at $z$ to $\boldmath{1}^{A}$. Write $\Theta (a)$
for the set of subsets of $A$ containing $a$, then we have%
\[
\frac{\partial \varphi }{\partial z_{a}}(z)=\sum_{i:a\in L_{i}}\frac{1}{%
z_{L_{i}}}=\sum_{S\in \Theta (a)}\frac{n_{S}}{z_{S}} 
\]%
so if $\Theta (a-b)$ is the set of coalitions containing $a$ and not $b$,
and $S^{c}$ is $A\diagdown S$, we have%
\begin{equation}
\frac{\partial \varphi }{\partial z_{a}}(z)=\frac{\partial \varphi }{%
\partial z_{b}}(z)\Longleftrightarrow \sum_{S\in \Theta (a-b)}(\frac{n_{S}}{%
z_{S}}-\frac{n_{S^{c}}}{z_{S^{c}}})=0  \label{16}
\end{equation}%
Thus $f^{nsh}(u)=z$ holds iff the right hand equation above holds for $|A|-1$
independent pairs $a,b$.

\paragraph{Constructing an example violating $SP^{+}$}

We note first that $SP^{+}$ is equivalent to the following property: if at
profile $u$ a set of $K$ agents have identical preferences $S$, and $%
u^{\prime }$ is obtained from $u$ when they all report instead $S\cup T$,
ceteris paribus, then the total weight of $S$ decreases weakly from $u$ to $%
u^{\prime }$. Indeed by $SP^{+}$, when our $K$ agents misreport their
preferences one at a time, the weight of $a$ must decrease weakly at each
step.

We fix now $A=\{a,b,c_{1},c_{2}\}$ and two mixtures $z,z^{\prime }\in \Delta
(A)$, both symmetric in $c_{1},c_{2}$ and such that%
\begin{equation}
0<z_{a}<z_{a}^{\prime }\text{ , }0<z_{b}<z_{b}^{\prime }\text{ , }%
z_{c}>z_{c}^{\prime }>0  \label{17}
\end{equation}%
(where $z_{c}$ stands for $z_{c_{1}}=z_{c_{2}}$). We show that, under some
additional restrictions on $z,z^{\prime }$ we can choose $u\in \Omega $ and
an integer $K$ such that $f^{nsh}(u)=z$ and $f^{nsh}(u^{\prime })=z^{\prime
} $, where $u^{\prime }$ is obtained from $u$ when $K$ agents of type $a$
switch to $ab$. This contradicts $SP^{+}$.

The profile $u$ is symmetric in $c_{1},c_{2}$ and\ we write $n_{ac}$ in lieu
of $n_{ac_{1}}$ etc..Then $\varphi (u)=z$ holds iff the two equations (\ref%
{16}) applied respectively to $a$, $c_{1}$ and to $b$, $c_{1}$, are true:
note that (\ref{16}) for $c_{1},c_{2}$ is obtained by the symmetry
assumption. These two equations are%
\begin{equation}
(\frac{n_{a}}{z_{a}}-\frac{n_{bcc}}{z_{bcc}})+(\frac{n_{ab}}{z_{ab}}-\frac{%
n_{cc}}{z_{cc}})+(\frac{n_{ac_{2}}}{z_{ac_{2}}}-\frac{n_{bc_{1}}}{z_{bc_{1}}}%
)+(\frac{n_{abc_{2}}}{z_{abc_{2}}}-\frac{n_{c_{1}}}{z_{c_{1}}})=0  \label{18}
\end{equation}%
\begin{equation}
(\frac{n_{b}}{z_{b}}-\frac{n_{acc}}{z_{acc}})+(\frac{n_{ab}}{z_{ab}}-\frac{%
n_{cc}}{z_{cc}})+(\frac{n_{bc_{2}}}{z_{bc_{2}}}-\frac{n_{ac_{1}}}{z_{ac_{1}}}%
)+(\frac{n_{abc_{2}}}{z_{abc_{2}}}-\frac{n_{c_{1}}}{z_{c_{1}}})=0  \label{19}
\end{equation}

We choose $u$ in such a way that all parenthesis above are null, that is we
pick five positive parameters $\alpha ,\beta ,\gamma ,\delta ,\varepsilon $,
such that%
\begin{equation}
\frac{n_{a}}{z_{a}}=\frac{n_{bcc}}{z_{bcc}}=\alpha K\text{ ; }\frac{n_{b}}{%
z_{b}}=\frac{n_{acc}}{z_{acc}}=\beta K\text{ ; }\frac{n_{c}}{z_{c}}=\frac{%
n_{abc}}{z_{abc}}=\gamma K  \label{20}
\end{equation}%
\begin{equation}
\frac{n_{ab}}{z_{ab}}=\frac{n_{cc}}{z_{cc}}=\delta K\text{ ; }\frac{n_{ac}}{%
z_{ac}}=\frac{n_{bc}}{z_{bc}}=\varepsilon K  \label{21}
\end{equation}

Note that we must have $n_{a}\geq K\Longleftrightarrow \alpha \geq \frac{1}{%
z_{a}}$ in order to construct $u^{\prime }$ by transforming $K$ agents who
only like $\{a\}$ to agents who like $a$ and $b$. And if the coordinates of $%
z$ and the numbers $\alpha ,\cdots ,\varepsilon $ are all rational, we can
choose $K$ large enough so that the above system delivers integers $n_{S}$
for all $S$.

The profile $u^{\prime }$ has $n_{a}^{\prime }=n_{a}-K$ and $n_{ab}^{\prime
}=n_{ab}+K$, and other terms $n_{S}$ are as in $u$. The desired equality $%
f^{nsh}(u^{\prime })=z^{\prime }$ requires two equations like (\ref{18}) and
(\ref{19}). For instance (\ref{18}) becomes%
\[
(\frac{n_{a}}{z_{a}^{\prime }}-\frac{n_{bcc}}{z_{bcc}^{\prime }})+(\frac{%
n_{ab}}{z_{ab}^{\prime }}-\frac{n_{cc}}{z_{cc}^{\prime }})+(\frac{n_{ac}}{%
z_{ac}^{\prime }}-\frac{n_{bc}}{z_{bc}^{\prime }})+(\frac{n_{abc}}{%
z_{abc}^{\prime }}-\frac{n_{c}}{z_{c}^{\prime }})=K(\frac{1}{z_{a}^{\prime }}%
-\frac{1}{z_{ab}^{\prime }}) . 
\]%
Taking (\ref{20}), (\ref{21}) into account this becomes%
\[
(\frac{z_{a}}{z_{a}^{\prime }}-\frac{z_{bcc}}{z_{bcc}^{\prime }})\alpha +(%
\frac{z_{ab}}{z_{ab}^{\prime }}-\frac{z_{cc}}{z_{cc}^{\prime }})\delta +(%
\frac{z_{ac}}{z_{ac}^{\prime }}-\frac{z_{bc}}{z_{bc}^{\prime }})\varepsilon
+(\frac{z_{abc}}{z_{abc}^{\prime }}-\frac{z_{c}}{z_{c}^{\prime }})\gamma =%
\frac{1}{z_{a}^{\prime }}-\frac{1}{z_{ab}^{\prime }} . 
\]%
\[
\Longleftrightarrow \frac{z_{a}-z_{a}^{\prime }}{z_{a}^{\prime }\cdot
(1-z_{a}^{\prime })}\cdot \alpha +\frac{z_{ab}-z_{ab}^{\prime }}{%
z_{ab}^{\prime }\cdot (1-z_{ab}^{\prime })}\cdot \delta +\frac{%
z_{ac}-z_{ac}^{\prime }}{z_{ac}^{\prime }\cdot (1-z_{ac}^{\prime })}\cdot
\varepsilon +\frac{z_{c}^{\prime }-z_{c}}{z_{c}^{\prime }\cdot
(1-z_{c}^{\prime })}\cdot \gamma =\frac{z_{b}^{\prime }}{z_{a}^{\prime
}\cdot z_{ab}^{\prime }}. 
\]%
Now we use inequalities (\ref{17}) to check that in the above sum, all
numerators except $z_{ac}-z_{ac}^{\prime }$ are negative. Therefore $%
z_{ac}-z_{ac}^{\prime }$ \ is positive and we can rewrite this equation as%
\begin{equation}
\frac{z_{ac}-z_{ac}^{\prime }}{z_{ac}^{\prime }\cdot (1-z_{ac}^{\prime })}%
\cdot \varepsilon =\frac{z_{b}^{\prime }}{z_{a}^{\prime }\cdot
z_{ab}^{\prime }}+\frac{z_{c}-z_{c}^{\prime }}{z_{c}^{\prime }\cdot
(1-z_{c}^{\prime })}\cdot \gamma +\frac{z_{a}^{\prime }-z_{a}}{z_{a}^{\prime
}\cdot (1-z_{a}^{\prime })}\cdot \alpha +\frac{z_{ab}^{\prime }-z_{ab}}{%
z_{ab}^{\prime }\cdot (1-z_{ab}^{\prime })}\cdot \delta  \label{22}
\end{equation}%
where all fractions are positive on both sides.

The second equation we need to ensure $f^{nsh}(u^{\prime })=z^{\prime }$ is
the counterpart of (\ref{19}) and reads%
\[
(\frac{n_{b}}{z_{b}^{\prime }}-\frac{n_{acc}}{z_{acc}^{\prime }})+(\frac{%
n_{ab}}{z_{ab}^{\prime }}-\frac{n_{cc}}{z_{cc}^{\prime }})+(\frac{n_{bc}}{%
z_{bc}^{\prime }}-\frac{n_{ac}}{z_{ac}^{\prime }})+(\frac{n_{abc}}{%
z_{abc}^{\prime }}-\frac{n_{c}}{z_{c}^{\prime }})+\frac{K}{z_{ab}^{\prime }}%
=0. 
\]

A similar computation using (\ref{20}), (\ref{21}) to change the terms $%
n_{S} $ into $z_{S}$ and inequalities (\ref{17}) to sign the fractions gives%
\begin{equation}
\frac{1}{z_{ab}^{\prime }}=\frac{z_{b}^{\prime }-z_{b}}{z_{b}^{\prime }\cdot
(1-z_{b}^{\prime })}\cdot \beta +\frac{z_{ab}^{\prime }-z_{ab}}{%
z_{ab}^{\prime }\cdot (1-z_{ab}^{\prime })}\cdot \delta +\frac{%
z_{ac}-z_{ac}^{\prime }}{z_{ac}^{\prime }\cdot (1-z_{ac}^{\prime })}\cdot
\varepsilon +\frac{z_{c}-z_{c}^{\prime }}{z_{c}^{\prime }\cdot
(1-z_{c}^{\prime })}\cdot \gamma  \label{23}
\end{equation}%
where again all ratios are positive.

We must show that the non negative rational numbers $\alpha ,\cdots
,\varepsilon $ can be chosen solving system (\ref{22}), (\ref{23}) and $%
\alpha \geq \frac{1}{z_{a}}$. Note that (\ref{22}) implies%
\[
\frac{z_{ac}-z_{ac}^{\prime }}{z_{ac}^{\prime }\cdot (1-z_{ac}^{\prime })}%
\cdot \varepsilon >\frac{z_{b}^{\prime }}{z_{a}^{\prime }\cdot
z_{ab}^{\prime }}+\frac{z_{a}^{\prime }-z_{a}}{z_{a}^{\prime }\cdot
(1-z_{a}^{\prime })}\cdot \frac{1}{z_{a}} 
\]%
and (\ref{23}) gives%
\[
\frac{z_{ac}-z_{ac}^{\prime }}{z_{ac}^{\prime }\cdot (1-z_{ac}^{\prime })}%
\cdot \varepsilon <\frac{1}{z_{ab}^{\prime }} 
\]

We can choose $\varepsilon $ meeting these two inequalities if and only if%
\[
\frac{z_{b}^{\prime }}{z_{a}^{\prime }\cdot z_{ab}^{\prime }}+\frac{%
z_{a}^{\prime }-z_{a}}{z_{a}^{\prime }\cdot (1-z_{a}^{\prime })}\cdot \frac{1%
}{z_{a}}<\frac{1}{z_{ab}^{\prime }}\Longleftrightarrow \frac{z_{a}^{\prime
}-z_{a}}{z_{a}\cdot (1-z_{a}^{\prime })}<\frac{z_{a}^{\prime }-z_{b}^{\prime
}}{z_{ab}^{\prime }} 
\]%
\begin{equation}
\Longleftrightarrow z_{a}>\frac{z_{a}^{\prime }+z_{b}^{\prime }}{%
2-z_{a}^{\prime }+z_{b}^{\prime }}.  \label{24}
\end{equation}%
and in this case we can also pick $\alpha \geq \frac{1}{z_{a}}$ as well as $%
\beta ,\gamma ,\delta $ solving (\ref{22}), (\ref{23}).

Summing up the requirements on $z,z^{\prime }$: we need inequalities (\ref%
{17}), (\ref{24}) as well as $z_{ac}^{\prime }<z_{ac}\Longleftrightarrow
z_{a}^{\prime }-z_{b}^{\prime }<z_{a}-z_{b}$. Note that (\ref{24}) and $%
z_{a}^{\prime }>z_{a}$ together imply $z_{a}^{\prime }>z_{b}^{\prime }$. We
can construct such a pair $z,z^{\prime }$ as follows.

Write $r$ the RHS in (\ref{24}), and check $r<z_{a}^{\prime }$ as long as so 
$z_{a}^{\prime }>z_{b}^{\prime }$. Thus it is enough to pick $z_{a}$ in the
interval $]\max \{r,z_{a}^{\prime }-z_{b}^{\prime }\},z_{a}^{\prime }[$, and
then to pick $z_{b}$ small enough that $z_{a}^{\prime }-z_{b}^{\prime
}<z_{a}-z_{b}$.

For instance we can choose%
\[
z_{a}=\frac{9}{20},\text{ }z_{b}=\frac{1}{20},\text{ }z_{c}=z_{c^{\prime }}=%
\frac{1}{4}\text{ \ ; \ }z_{a}^{\prime }=\frac{1}{2},\text{ }z_{b}^{\prime
}=z_{c}^{\prime }=z_{c^{\prime }}^{\prime }=\frac{1}{6} 
\]%
the system (\ref{22}), (\ref{23}) is then%
\[
\frac{3}{20}\varepsilon =\frac{1}{2}+\frac{3}{5}\gamma +\frac{3}{4}\delta +%
\frac{1}{5}\alpha 
\]%
\[
\frac{3}{5}\gamma +\frac{3}{20}\varepsilon +\frac{21}{25}\beta +\frac{3}{4}%
\delta =\frac{3}{2} 
\]%
where we recall the constraint $n_{a}\geq K\Leftrightarrow \frac{9}{20}%
\alpha \geq 1$.

A relatively simple solution of the system above is%
\[
\alpha =\frac{25}{11},\text{ }\gamma =\frac{5}{11},\text{ }\varepsilon =%
\frac{90}{11},\text{ }\beta =\delta =0 
\]%
for which we derive the profile $u\in \mathcal{\Omega }$ by system (\ref{20}%
), (\ref{21}). Here $K=44$ is the smallest integer delivering integer
coordinates, and we end up with 860 agents and the profile%
\[
n_{a}=45,\text{ }n_{bcc^{\prime }}=55\text{ ;\ }n_{c}=n_{c^{\prime }}=5,%
\text{ }n_{abc}=n_{abc^{\prime }}=15\text{ ;\ }n_{ac}=n_{ac^{\prime }}=252; 
\]
\[
\text{ }n_{bc}=n_{bc^{\prime }}=108. 
\]

\paragraph{Example where the NMP rule violates (a slightly stronger version
of) $SP^{0}$}

We use the same technique. Set $A=\{a,a^{\prime },a^{\prime \prime },b,c\}$
and construct two profiles $u,u^{\prime }$ such that $u$ is obtained from $%
u^{\prime }$ when $K$ agents who all like $\{a,a^{\prime },a^{\prime \prime
},b\}$ all declare $\{a,a^{\prime },a^{\prime \prime }\}$ and end up better
off even though they cannot consume $b$ anymore (so $u^{\prime }$ is the
true profile). This property implies a group version of $SP^{0}$.

At profile $u$ the $K$ agents in question declare $\{a,a^{\prime },a^{\prime
\prime }\}$ and $\varphi (u)=z$; at $u^{\prime }$ they switch to $%
\{a,a^{\prime },a^{\prime \prime },b\}$ (nothing else changes) and $\varphi
(u^{\prime })=z^{\prime }$. The profiles are entirely symmetric in $%
a,a^{\prime },a^{\prime \prime }$. We define%
\[
z_{a}=\frac{1}{6},z_{b}=\frac{1}{32},z_{c}=\frac{15}{32}\text{ ; }%
z_{a}^{\prime }=\frac{1}{16},z_{b}^{\prime }=\frac{1}{4},z_{c}^{\prime }=%
\frac{9}{16} 
\]%
Note that $z_{aaab}^{\prime }<z_{aaa}$ as desired but $z_{aab}<z_{aab}^{%
\prime }$.

We have six types of preferences and ten homogenous coalitions of which the
sizes meet the analog of system (\ref{12}), (\ref{13}) for some positive
parameters $\gamma ,\delta $:%
\[
\frac{n_{aaa}}{z_{aaa}}=\frac{n_{bc}}{z_{bc}}=K\text{ , }\frac{n_{aaab}}{%
z_{aaab}}=\frac{n_{c}}{z_{c}}=\gamma K\text{ , }\frac{n_{ac}}{z_{ac}}=\frac{%
n_{aab}}{z_{aab}}=\delta K 
\]%
(we have three coalitions who like one of the $a$-s and $c$, and another
three who like two of the $a$-s and $b$). This implies $\varphi (u)=z$. To
ensure $\varphi (u^{\prime })=z^{\prime }$ we need two instances of the
first order condition (\ref{16}), respectively for $b,c$ and for $a^{\ast
},b $ where $a^{\ast }$ is an arbitrary selection from $a,a^{\prime
},a^{\prime \prime }$; symmetry implies (\ref{16}) between two $a$-s.

Straightforward computations as above, omitted for brevity, show that (\ref%
{16}) for $b$,$c$ reduces to%
\[
\frac{z_{aaab}-z_{aaab}^{\prime }}{z_{aaab}^{\prime }\cdot z_{c}^{\prime }}%
\gamma K+\frac{16}{7}K=3\frac{z_{ac}-z_{ac}^{\prime }}{z_{ac}^{\prime }\cdot
z_{aab}^{\prime }}\delta K 
\]%
and for $a^{\ast },b$%
\[
\frac{z_{aaa}-z_{aaa}^{\prime }}{z_{aaa}^{\prime }\cdot z_{bc}^{\prime }}K+%
\frac{z_{a^{\ast }c}-z_{a^{\ast }c}^{\prime }}{z_{a^{\ast }c}^{\prime }\cdot
z_{aab}^{\prime }}\delta K=\frac{16}{3}K 
\]%
With our choice of $z,z^{\prime }$ these equations boil down to%
\[
\frac{32}{85}\gamma +\frac{16}{7}=3(\frac{2}{45}\delta )\text{ ; }\frac{160}{%
39}+\frac{2}{45}\delta =\frac{16}{3} 
\]%
\[
\Longrightarrow \gamma =\frac{340}{91}\text{ , }\delta =\frac{360}{13}. 
\]%
Finally we pick $K=8\times 19\times 91=13832$ and get%
\begin{eqnarray*}
n_{aaa} &=&n_{bc}=K\text{ ; }n_{aaab}=17\times 85\times 19=27455\text{ ; }%
n_{c}=15\times 85\times 19=24225 \\
\text{ }n_{aab} &=&35\times 90\times 56\simeq 176400,n_{ac}=61\times
90\times 56\simeq 307440.
\end{eqnarray*}

\end{document}